\documentclass[a4paper,usenatbib]{mn2e}

\usepackage{amsmath}	% Advanced maths commands
\usepackage[T1]{fontenc}
\usepackage{ae,aecompl}

\usepackage{graphicx}	% Including figure files
\usepackage{amssymb}	% Extra maths symbols

\usepackage{wrapfig}
\usepackage{lscape}

\usepackage{hyperref}	% Hyperlinks
\hypersetup{colorlinks=true,linkcolor=blue,citecolor=blue,filecolor=blue,urlcolor=blue}
\newcommand{\mo}{{\rm M}_\odot}
\newcommand{\mz}{$\langle {\rm Z} \rangle$}
\newcommand{\mvz}{$\langle {\rm V_{z}} \rangle$}

\title[Warps and waves]{Warps and waves in the stellar discs of the Auriga cosmological simulations}

\author[F. A. Gomez et al.]{Facundo A. G\'omez$^{1}$\thanks{E-mail: fgomez@mpa-garching.mpg.de}, Simon D. M. White$^{1}$,
Robert J. J. Grand$^{2,3}$, \newauthor Federico Marinacci$^{4}$,
Volker Springel$^{2,3}$ and R{\"u}diger Pakmor$^{2}$\\
$^{1}$Max-Planck-Institut f\"ur Astrophysik, Karl-Schwarzschild-Str. 1, D-85748, Garching, Germany\\
$^{2}$Heidelberger Institut f\"ur Theoretische Studien, Schloss-Wolfsbrunnenweg 35, 69118 Heidelberg, Germany\\
$^{3}$Zentrum f\"ur Astronomie der Universitat Heidelberg, Astronomisches Recheninstitut, Monchhofstr. 12-14, 69120 Heidelberg, Germany\\
$^{4}$Department of Physics, Kavli Institute for Astrophysics and Space Research, MIT, Cambridge, MA 02139, USA\\
}

\date{Accepted XXX. Received YYY; in original form ZZZ}

\pubyear{2015}

\begin{document}
\label{firstpage}
\pagerange{\pageref{firstpage}--\pageref{lastpage}}
\maketitle

% Abstract of the paper
\begin{abstract}
Recent studies have revealed an oscillating asymmetry in the vertical
structure of the Milky Way's disc. Here we analyze 16 high-resolution,
fully cosmological simulations of the evolution of individual Milky
Way-sized galaxies, carried out with the MHD code AREPO.  At redshift
zero, about $70\%$ of our galactic discs show strong vertical
patterns, with amplitudes that can exceed 2 kpc.  Half of these are
typical `integral sign' warps. The rest are oscillations similar to
those observed in the Milky Way. Such structures are thus expected to
be common. The associated mean vertical motions can be as large as
30~km/s.  Cold disc gas typically follows the vertical patterns seen
in the stars.  These perturbations have a variety of causes: close
encounters with satellites, distant flybys of massive objects,
accretion of misaligned cold gas from halo infall or from mergers.
Tidally induced vertical patterns can be identified in both young and
old stellar populations, whereas those originating from cold gas
accretion are seen mainly in the younger populations. Galaxies with
regular or at most weakly perturbed discs are usually, but not always,
free from recent interactions with massive companions, although we
have one case where an equilibrium compact disc reforms after a
merger.

\end{abstract}

% Select between one and six entries from the list of approved keywords.
% Don't make up new ones.

%%%%%%%%%%%%%%%%%%%%%%%%%%%%%%%%%%%%%%%%%%%%%%%%%%

%%%%%%%%%%%%%%%%% BODY OF PAPER %%%%%%%%%%%%%%%%%%

\section{Introduction}
\label{sec:intro}

Vertical perturbations in the outer regions of galactic discs were first
discovered through surveys of HI 21cm lines \citep[][ and references
therein]{1992ARA&A..30...51B}. Follow up studies based on large
observational samples of stellar discs
revealed that many of them also show perturbed vertical
structures \citep[e.g.][]{1998A&A...337....9R,2006NewA...11..293A, 2016MNRAS.461.4233R}. The
most common morphology is what is known as an S-shaped or `integral
sign' warp, but other kinds of distortion are also present.

Broadly speaking, two major mechanisms produce vertical perturbations
of the otherwise flat outer discs of spirals
\citep{2013pss5.book..923S}.  The first is misaligned accretion of
high angular momentum cold gas. Such accretion can result from a close
encounter with a gas-rich satellite \citep[e.g. the Magellanic Stream,
  see][]{2008A&ARv..15..189S} or from misaligned infall from the
cosmic web or from a cooling hot gas halo
\citep[e.g.][]{1999MNRAS.303L...7J, 2010MNRAS.408..783R, 2013MNRAS.434.3142A}. Such
accretion is expected to produce vertical patterns which are most
evident in the outer disc gas and in young stars formed from it. The
second mechanism is tidal distortion of a pre-existing disc by an
external perturber, such as a low-mass (perhaps merging) satellite
with a small pericentre \citep[e.g.][]{1989MNRAS.237..785O,
  1993ApJ...403...74Q, 1999MNRAS.304..254V, 2003ApJ...583L..79B,
  2009ApJ...700.1896K, 2013MNRAS.429..159G, 2015arXiv151101503D} a
more massive satellite on a distant fly-by
\citep[][]{2000ApJ...534..598V, 2014ApJ...789...90K, G16}, or a
misaligned outer dark matter halo \citep[e.g.][]{1999ApJ...513L.107D,
  1999MNRAS.303L...7J, 2006MNRAS.370....2S}.

Several studies have indicated that our own Galactic stellar disc has
been vertically perturbed
\citep[e.g.][]{2002A&A...394..883L,2006A&A...451..515M}. However, the
morphology of the vertical structure appears more complex than a
traditional S- or U-shaped warp. Rather, an oscillating asymmetry,
i.e. a global bending pattern, is suggested, with an amplitude that grows
as a function of galactocentric distance.  The first evidence for such
a bending pattern was presented by \citet{2012ApJ...750L..41W}, who
identified a North-South asymmetry in the distribution of a sample of
main sequence stars in the Solar Neighborhood ($R_{\rm helio} \lesssim
1$ kpc and $|Z| \lesssim 2$ kpc) \citep[see
  also][]{2013ApJ...777...91Y}. \citet{2015ApJ...801..105X} extended
this result by characterizing the vertical distribution of a sample of
main sequence stars, obtained from the Sloan Digital Sky Survey
\citep{sdss}, as a function of galactocentric distance. They found
radially extended oscillatory behaviour in the direction of the
Galactic anticentre \citep[see also][]{2015MNRAS.452..676P}. Similar
results were obtained by \citet{2014ApJ...791....9S}, who used a
sample of main sequence turn-off stars from Pan-STARRS1
\citep{2010SPIE.7733E..0EK} to map the anticentre stellar distribution
to a heliocentric distance of $\sim 17$ kpc.  This study revealed that
the Monoceros Ring \citep{new02,yanny03}, a large and complex stellar
structure in the outer Milky Way disc, exhibits a North-South
asymmetry with the southern and northern parts dominating the regions
closer and further from the Sun, respectively.  More recently,
\citet{2016arXiv160407501M} extended this analysis by creating
three-dimensional maps of the Monoceros Ring, concluding that it is
well described by two concentric circles alternately seen in the South
and the North, suggestive of a wave propagating away from a common
origin. It is important to note that the systematics of such vertical
oscillations in external galaxies are unknown. Most studies attempting
to characterize the vertical structure of discs have focused on
edge-on systems in which oscillations are difficult to detect because
of projection effects.

The idea that the Monoceros Ring reflects a vertical oscillation of
the Galactic disc has been supported by many numerical studies
\citep[e.g.][]{2008ApJ...688..254K,2008ApJ...676L..21Y,2009MNRAS.397.1599Q,2011Natur.477..301P,2013MNRAS.429..159G,2014MNRAS.440.1971W,2015MNRAS.454..933D}.
However, most of these have been based on cosmologically motivated but
idealized simulations. \citet[][hereafter G16]{G16} presented the
first model of a Monoceros-like ring obtained from a fully
cosmological simulation of the formation of a Milky Way-mass
galaxy. This model reproduces, qualitatively, many of the observed
properties of the Monoceros Ring. However, the study included a single
simulation, and so could not characterize the expected frequency of
such structures in the galaxy population. Here, we address this point
by analyzing a suite of 16 high-resolution magnetohydrodynamical
simulations of the formation and evolution of a Milky Way-mass galaxy
\citep[the Auriga Project,][hereafter GR16]{2016arXiv161001159G}.  In
Section~\ref{sec:simu} we briefly introduce the simulations and
discuss their main properties. We characterize the present-day
vertical structure of their discs in Section~\ref{sec:pres-day}. In
Section~\ref{sec:hists} we connect this vertical structure with the
recent evolutionary history of the individual galaxies. We summarize
and discuss our results in Section~\ref{sec:summ}.

\section{Simulations}
\label{sec:simu}

\begin{table*}
\centering
\caption{Table of simulation parameters. The columns are 1) Simulation
  Id code used in this work 2) Model
  name in the Auriga Project reference paper (GR16); 3) Virial mass; 4) Virial radius; 5) Stellar mass; 6) Disc
  stellar mass; 7) Disc radial scale length; 8) Bulge stellar mass; 9)
  Bulge effective radius; 10) Sersic index of the bulge, and 11) Disc to
  total mass ratio. See GR16 for definitions }
\label{t1}
\begin{tabular}{c c c c c c c c c c c}
\hline
Id & Run & $M_{\rm vir}$ $(\rm 10^{12} M_{\odot})$ & $R_{\rm vir}$ (kpc) & $M_{*}$ $(\rm 10^{10} M_{\odot})$ & $M_{\rm d}$ $(\rm 10^{10} M_{\odot})$ & $R_{\rm d}$ (kpc) & $M_{\rm b}$ $(\rm 10^{10} M_{\odot})$ & $R_{\rm eff}$ (kpc) & $n$ & $D/T$  \\
\hline
  S1   & Au-5  &   1.19  &  223.09    &   6.72     &  4.38     &  3.80     &  1.95     &  0.87     &  1.02      &  0.69\\
  S2   & Au-6  &   1.04  &  213.82    &   4.75     &  3.92     &  4.53     &  0.67     &  1.30     &  1.01      &  0.85\\
  S3   & Au-12  &    1.09 &    217.12   &    6.01    &   4.33    &   4.03    &   1.48    &   1.05    &   1.07     &   0.75\\
  S4   & Au-17  &    1.03 &    212.77   &    7.61    &   2.67    &   2.82    &   4.14    &   1.11     &  0.71      &  0.39\\
  S5   & Au-18  &    1.22 &    225.29   &    8.04    &   5.17    &   3.03    &   1.98    &   1.06     &  0.79      &  0.72\\
  S6   &  Au-2  &   1.91  &   261.76   &    7.05   &    4.63   &    5.84   &    1.45   &    1.34   &    0.99    &    0.76 \\
  S7   & Au-3  &   1.46  &   239.02   &    7.75   &    6.29   &    7.50   &    2.10   &    1.51   &    1.06    &    0.75 \\
  S8   & Au-9  &   1.05  &   214.22   &    6.10    &   3.57    &   3.05    &   2.03    &   0.94    &   0.84     &   0.64\\
  S9   & Au-15  &    1.22 &    225.40   &    3.93    &   3.14    &   4.00    &   0.39    &   0.90    &   1.02     &   0.89\\
  S10 & Au-23  &    1.58 &    245.27   &    9.02    &   6.17    &   4.03    &   2.42    &   1.26     &  0.94      &  0.73\\
  S11 & Au-24  &    1.49 &    240.86   &    6.55    &   3.68    &   5.40    &   2.18    &   0.93     &  0.9      &  0.63\\
  S12 & Au-16  &    1.50 &    241.48   &    5.41    &   4.77    &   7.84    &   1.00    &   1.56     &  1.18      &  0.83\\
  S13 & Au-19  &    1.21 &    224.57   &    5.32    &   3.88    &   4.31    &   1.02    &   1.02     &  1.13      &  0.79\\
  S14 & Au-21  &    1.45 &    238.64   &    7.72    &   5.86    &   4.93    &   1.48    &   1.36     &  1.29      &  0.80\\
  S15 & Au-25  &    1.22 &    225.30   &    3.14    &   2.59    &   6.30    &   0.76    &   2.44     &  1.69      &  0.77\\
  S16 & Au-27  &    1.75 &    253.81   &    9.61    &   7.21    &   4.21    &   1.70    &   0.92     &  1.00      &  0.81\\
\hline
\end{tabular}
\end{table*}

In this paper we analyze a suite of 16 fully cosmological magneto
hydrodynamical simulations of the formation of a Milky Way-like
galaxy. In the following we summarize the methodology and the main
characteristics of these simulations. A more detailed description can
be found in GR16.  

The simulations  were carried  out using the  $N$-body +  moving mesh,
magnetohydrodynamics  code \textrm{AREPO} \citep{2010MNRAS.401..791S}.
AREPO solves  the gravitational and collisionless dynamics  by using a
TreePM   approach  \citep{springel2005a}   and  discretizes   the  MHD
equations on a dynamic unstructured Voronoi mesh.  These equations are
solved  with a second  order Runge-Kutta  integration scheme  based on
high-accuracy  least-square spatial  gradient estimators  of primitive
variables \citep{2016MNRAS.455.1134P}, which  is an improvement on the
treatment      in     the      original      version     of      AREPO
\citep{2010MNRAS.401..791S}.     A   $\Lambda$CDM    cosmology,   with
parameters  $\Omega_{\rm  m} =  \Omega_{\rm  dm}  +  \Omega_{\rm b}  =
0.307$,  $\Omega_{\rm  b}  =  0.048$, $\Omega_{\Lambda}=  0.693$,  and
Hubble  constant $H_{0}  = 100~h$  km s$^{-1}$  Mpc$^{-1}$ =  67.77 km
s$^{-1}$ Mpc$^{-1}$, was adopted.  The haloes were first identified in
a lower  resolution simulation of a  periodic box of side  100 Mpc. To
obtain  a sample  of  relatively isolated  MW-size systems,  candidate
haloes were selected  within a narrow mass range  centered on $10^{12}
{\rm  M}_{\odot}$ and  required to  be at  least $1.37$  Mpc  from any
object  with  more than  half  their  mass.  Applying the  ``zoom-in''
technique, the galaxies were  re-simulated multiple times at different
resolution, increasing the mass resolution in each step by a factor of
8.  After  setting up  the initial dark  matter distribution,  gas was
added  by splitting  each original  dark matter  particle into  a dark
matter  particle--gas cell  pair.   The masses  assigned  to each  are
determined from  the cosmological baryon  mass fraction, and  they are
separated by  half the mean inter-particle  spacing. Their phase-space
coordinates are  chosen such that the  centre of mass  and velocity of
the pair  are those  of the original  dark matter particle.   Here, we
focus  on the  simulations  with  resolution level  4,  for which  the
typical  mass of a  high resolution  dark matter  particle is  $\sim 3
\times 10^{5}$ $\rm M_{\odot}$, and of  an initial gas cell is $\sim 4
\times  10^{4}$ $\rm M_{\odot}$.   The stellar  physical gravitational
softening length  grows with the scale  factor up to a  maximum of 369
pc,  after which it  is kept  constant.  The  softening length  of gas
cells  scales with the  mean radius  of the  cell, with  minimum value
equal to  the stellar  softening length at  all times.  Note  that, in
high density regions, gas cells are allowed to become smaller than the
gravitational  softening  length.   A  resolution  study  across  four
resolution levels was  presented in GR16.  This study  shows that many
galaxy   properties,  such  as   surface  density   profiles,  orbital
circularity  distributions  and  disc  vertical  structures  are  well
converged already at the resolution level used in this work.

The simulations include modeling of the critical physical processes
that govern galaxy formation, such as gas cooling/heating, star
formation, mass return and metal enrichment from stellar evolution,
the growth of supermassive black holes, and feedback from stellar
sources and from black hole accretion. The parameters that regulate
the efficiency of each process were chosen by comparing the results
obtained in simulations of cosmologically representative regions to a
wide range of observations of the galaxy population
\citep[][]{2013MNRAS.436.3031V,2014MNRAS.437.1750M}.  Magnetic fields
are implemented as described in \citet{2013MNRAS.432..176P}. A
homogeneous magnetic field is seeded at $z=127$, with strength
$10^{-14}$ Gauss oriented (arbitrarily) along one coordinate axis. The
choice of direction and strength has little effect on the evolution
\citep{2013MNRAS.432..176P,2015MNRAS.453.3999M}. From now on, we will
refer to these simulations as S$i$, with $i$ enumerating the
different initial conditions. In addition, throughout this work we will
compare our results with those presented in G16, who analyzed an
additional simulation performed with the same numerical set-up and
mass resolution. We will refer to this simulations as Aq-C4.

\begin{figure*}
\centering
\includegraphics[width=180mm,clip]{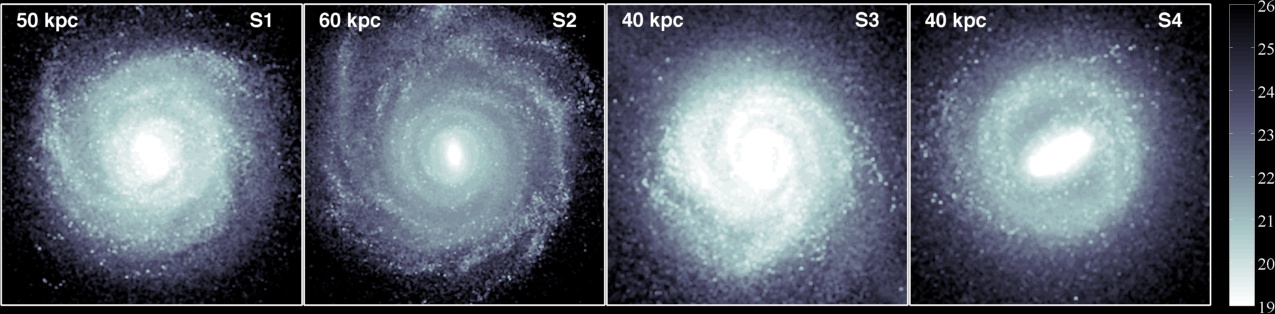}\\
\includegraphics[width=180mm,clip]{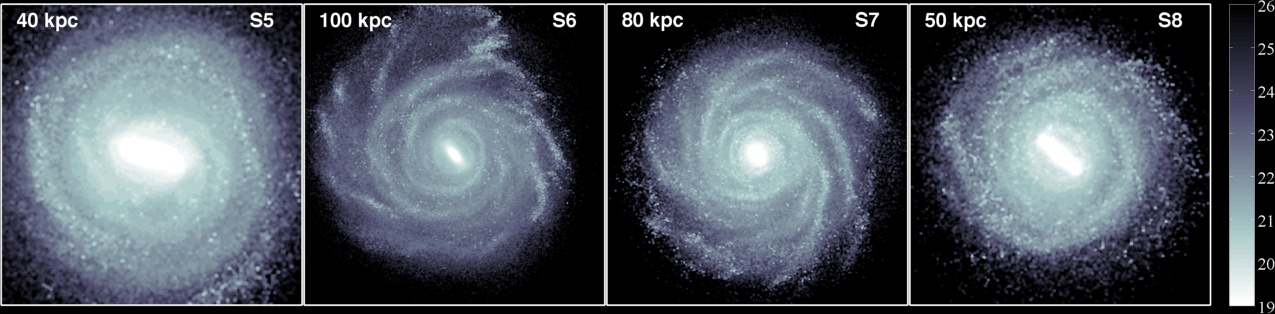}\\
\includegraphics[width=180mm,clip]{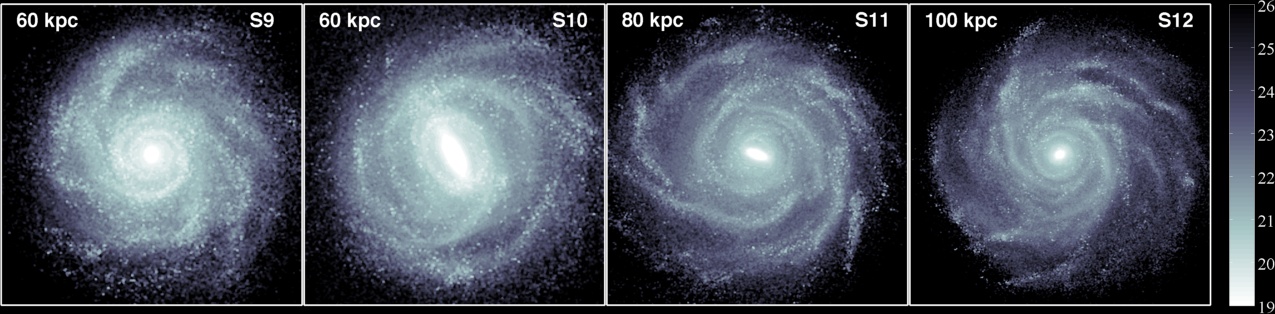}\\
\includegraphics[width=180mm,clip]{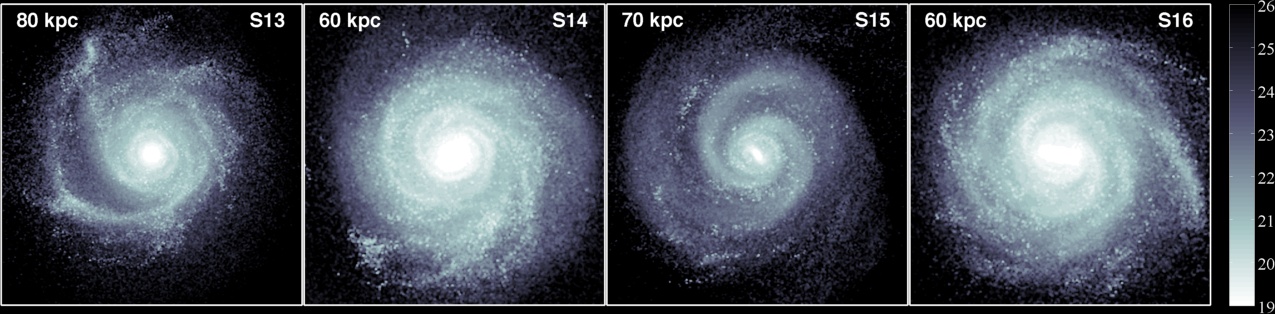}
\caption{Present-day face-on images of the $V-$band surface
  brightness, $\mu_{\rm v}$, of the galaxies analysed in this
  paper. The side length of each panel is indicated on its top left
  corner. Only particles that belong to the main host are
  considered. The colour bar indicates the scale for $\mu_{\rm v}$ in
  units of mag/arcsec$^2$.}
\label{fig:maps_sb}
\end{figure*}

Figure~\ref{fig:maps_sb} shows present-day $V-$band images of stellar
surface brightness for the 16 galaxies analysed in this work.  Only
star particles that at $z=0$ are gravitationally bound to the main
galaxy are used in making these images. In all cases, discs are shown
face-on. The side length of each panel is indicated in its top left
corner, and was chosen so such that the entire disc is visible.  As
discussed by GR16, these discs have a wide variety of morphologies and
extents. Some show multi-arm flocculent spiral structure (e.g. S6
and S12). Others have well defined two-armed spirals
(e.g. S15). Several have strong bars (S8 and S10) while others
have none (S9 and S13). The main properties of each simulated
galaxy are listed in Table~\ref{t1}. The disc/bulge decomposition is
made by simultaneously fitting of exponential and Sersic profiles to
the stellar surface density profiles. A detailed description of how
these parameters were obtained is given in \citep{Grand16}.

\begin{figure*}
\centering
\includegraphics[width=180mm,clip]{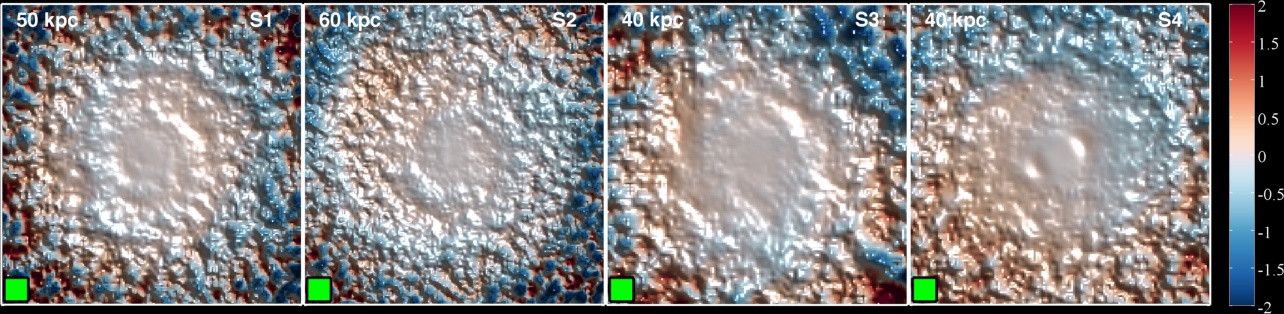}\\
\includegraphics[width=180mm,clip]{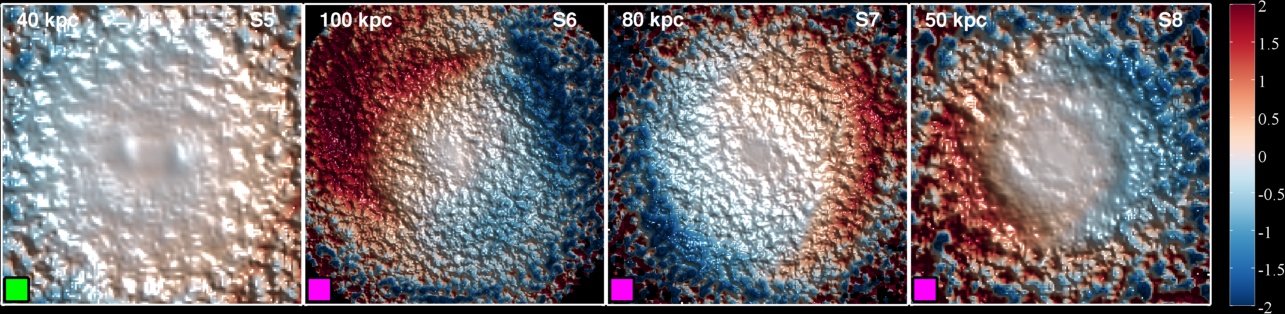}\\
\includegraphics[width=180mm,clip]{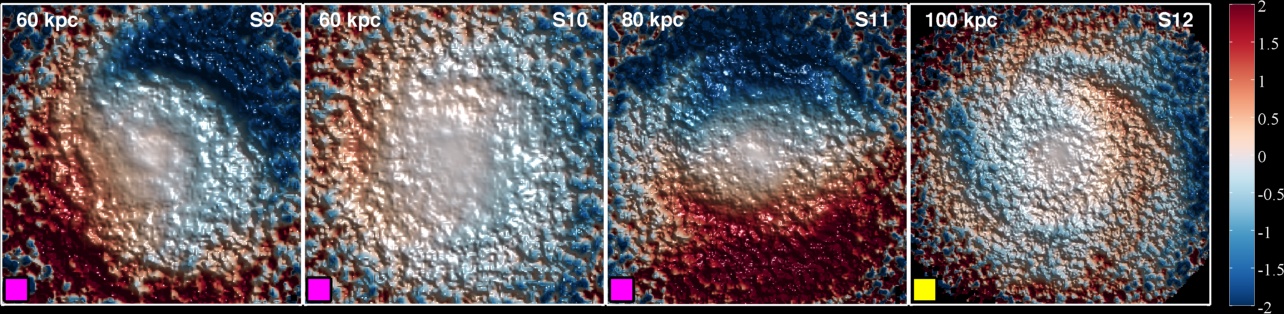}\\
\includegraphics[width=180mm,clip]{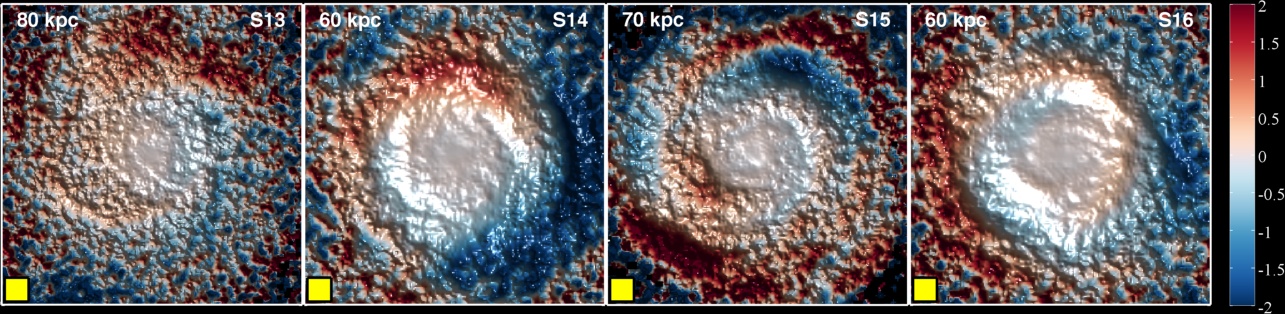}
\caption{Maps of the simulated stellar disc's mass-weighted mean
  height, \mz, at the present-day. The different map colours and
  relief indicate different values of \mz~in kpc. The box lengths were
  chosen to match those used in Figure~\ref{fig:maps_sb}. The
    colour coded symbols in the bottom left corner of each panel
    identify different type of vertical structure; green, magenta and
    yellow squares indicate weakly perturbed discs, simple warps, and
    more complex vertical patterns, respectively.}
\label{fig:maps_z}
\end{figure*}

\begin{figure*}
\centering
\includegraphics[width=180mm,clip]{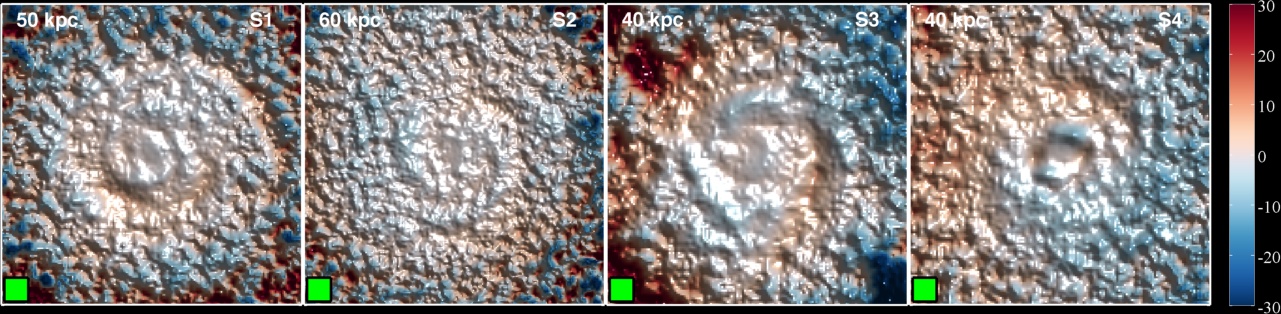}\\
\includegraphics[width=180mm,clip]{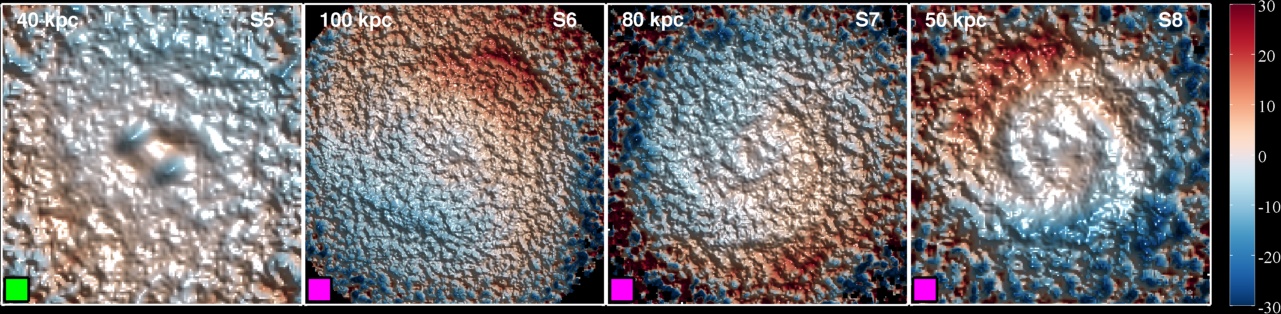}\\
\includegraphics[width=180mm,clip]{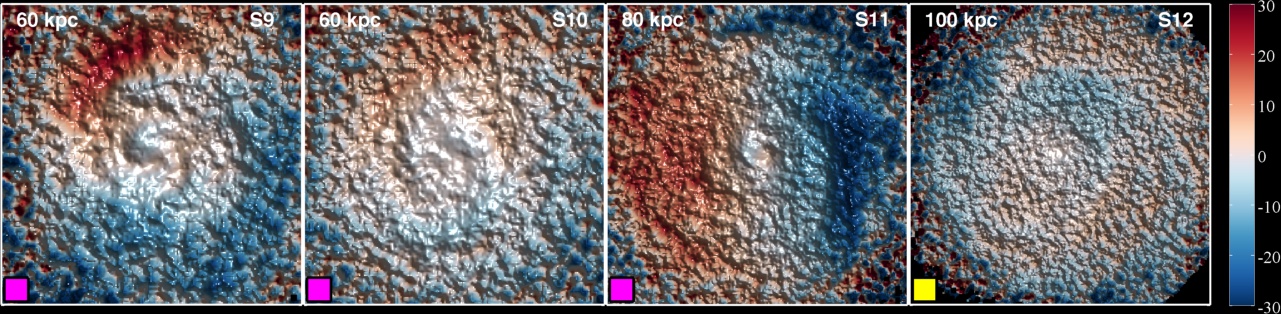}\\
\includegraphics[width=180mm,clip]{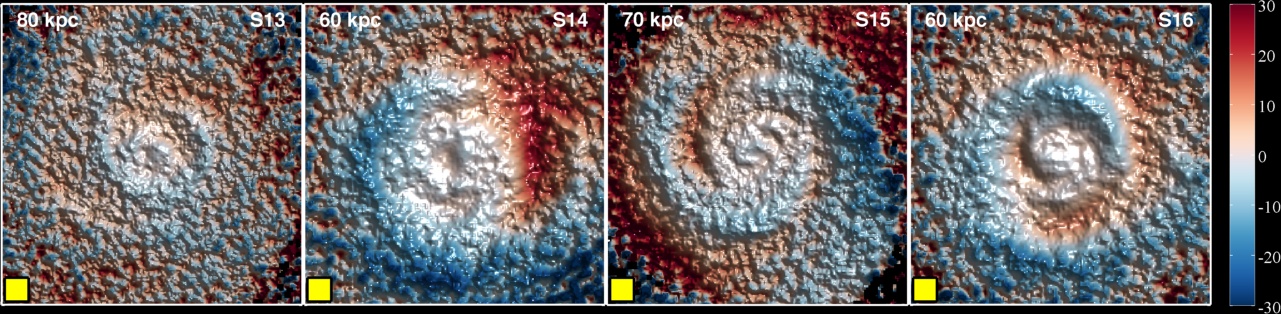}
\caption{As in Figure~\ref{fig:maps_z}, but for the mass-weighted mean
  vertical velocity, \mvz.  The different map colours and the relief
  indicate different values of \mvz~in km/s.}
\label{fig:maps_vz}
\end{figure*}

\begin{figure*}
\centering
\includegraphics[width=180mm,clip]{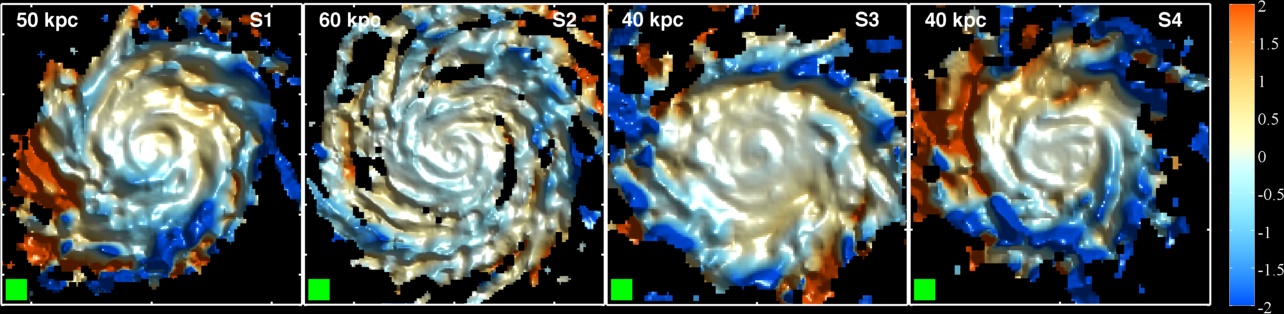}\\
\includegraphics[width=180mm,clip]{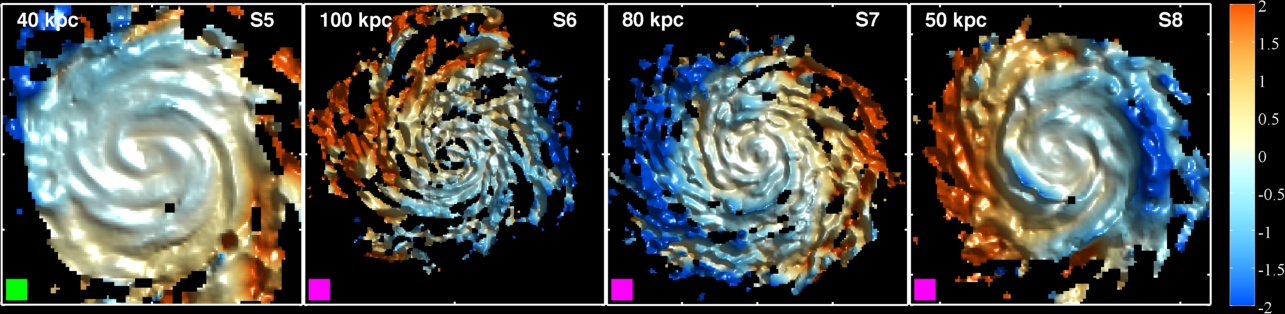}\\
\includegraphics[width=180mm,clip]{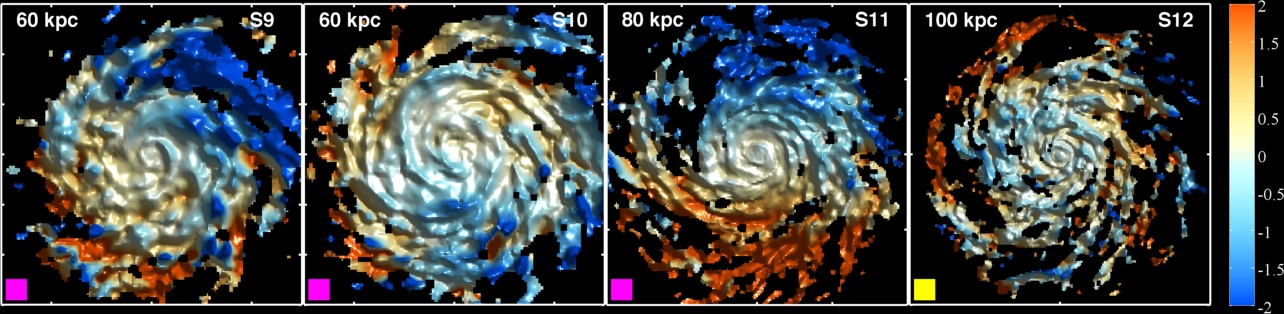}\\
\includegraphics[width=180mm,clip]{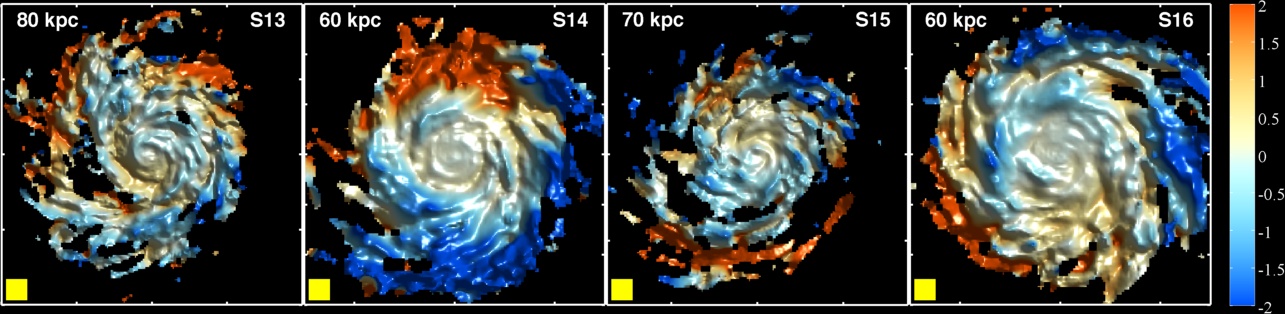}
\caption{As in Figure~\ref{fig:maps_z}, but for the cold star-forming
  gas. The different map colours and the relief indicate different values
  of \mz~in kpc.}
\label{fig:maps_gz}
\end{figure*}

\begin{figure*}
\centering
\includegraphics[width=180mm,clip]{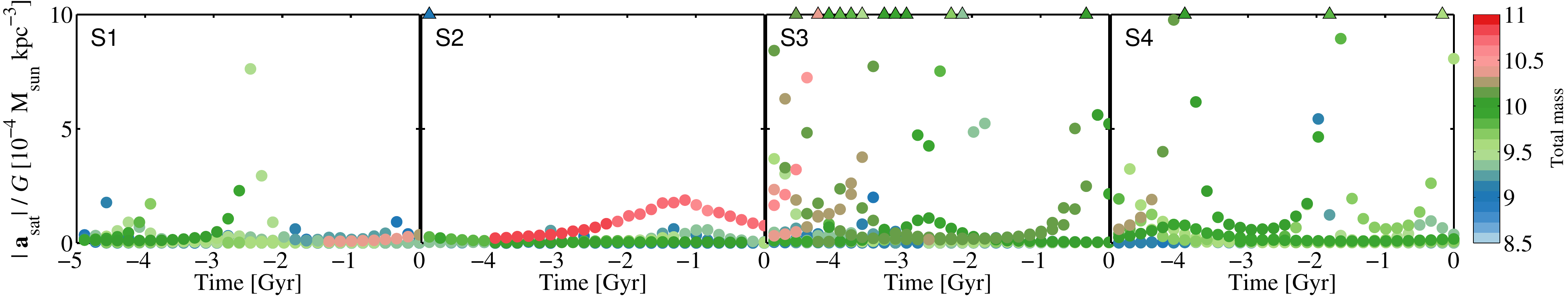}\\
\includegraphics[width=180mm,clip]{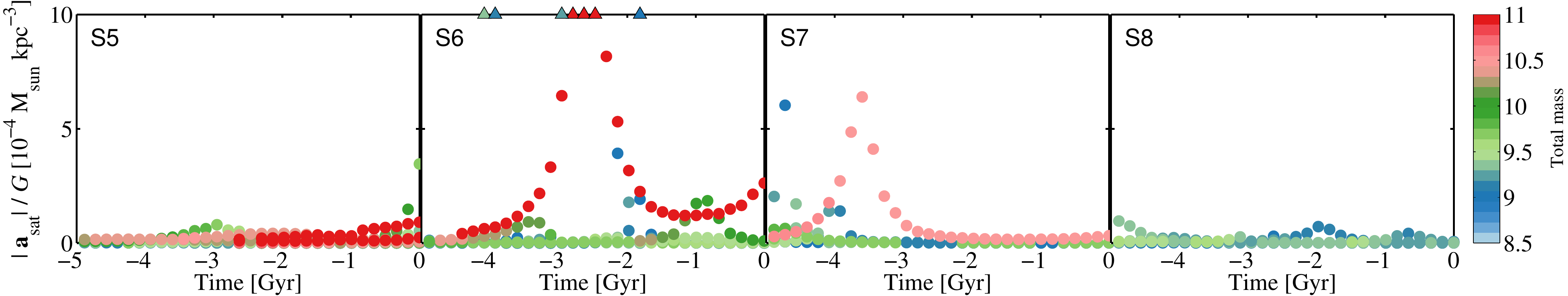}\\
\includegraphics[width=180mm,clip]{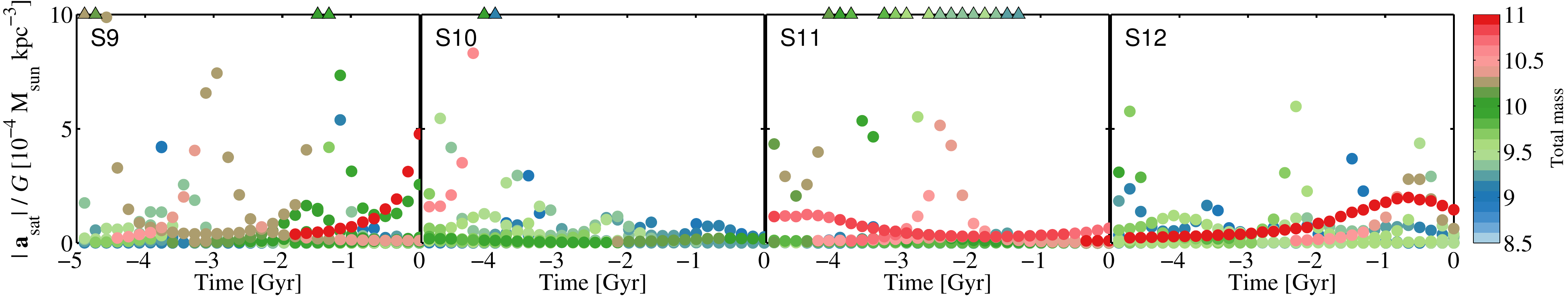}\\
\includegraphics[width=180mm,clip]{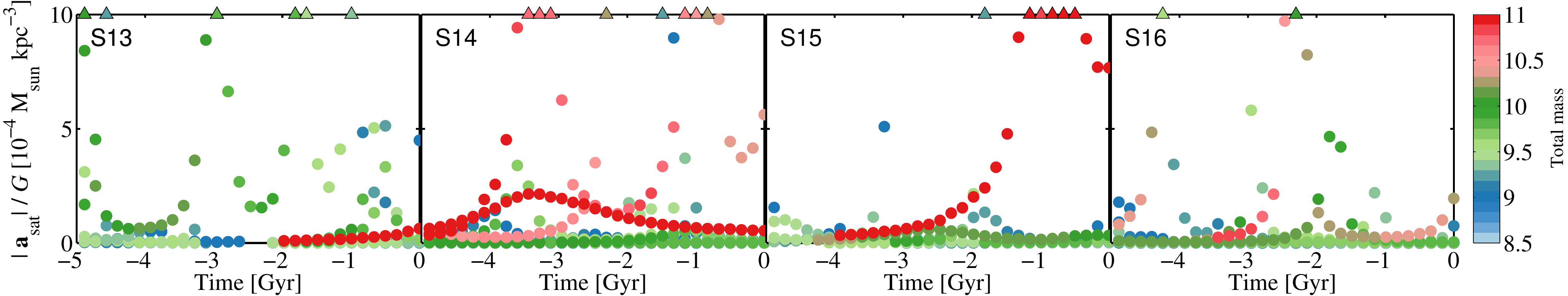}
\caption{The tidal field exerted on each host by its 20 most massive
  satellites as a function of time.  Points are colour -coded
  according to the mass of each satellite at the corresponding
  time.  Triangles indicate tidal field values that are above
    the Y-axis limit.}
\label{fig:tidal}
\end{figure*}

\section{Present-day discs vertical structure}
\label{sec:pres-day}

Following \citet{G16}, the discs were aligned with the X-Y plane by
iteratively computing, and aligning with the Z-direction, the total
angular momentum of the star particles within cylinders of 7 kpc
radius and of decreasing height. The star particles used in this
procedure were chosen to be less than $5$ Gyr old. To obtain a map of
the mass-weighted mean height, \mz, 
a regular Cartesian mesh of cell size 0.5 kpc aligned with the X-Y
plane was defined. On each node we centre a 1 kpc radius cylinder of $\pm 5$ kpc
height and compute the stellar mass-weighted \mz. It is important to notice
that these maps are weighted by mass and not by luminosity.  Thus, the
patterns seen in the vertical distribution do not necessarily follow
the younger and brighter stellar populations.

Figure~\ref{fig:maps_z} shows the \mz~maps for each galaxy. The panel
sides match those in Figure~\ref{fig:maps_sb}. Interestingly, the
sample of simulated discs exhibits a wide variety of vertical
structures. Some are almost flat (e.g. S1, S2, S3, S4 and
S5) comprising $\sim 30\%$ of the sample, while others (e.g. S6,
S7, S8, S9, S10 and S11) show strong $m=1$ patterns or
warps. The amplitude of the warps can be as large as \mz$~\gtrsim 2$
kpc in the outer regions (e.g. S6, S9 and S11). The warp angles
$\alpha$ of these five warped galaxies, defined as the angle between
the plane of the inner unwarped disc and the line connecting the
centre to the outermost tip of the warp, take values $\sim
2.5^{\circ}$, $2.7^{\circ}$, $4^{\circ}$, $7.4^{\circ}$, $2.3^{\circ}$
and $5^{\circ}$, respectively.  This is consistent with the range of
$\alpha$-values found observationally from optical data
\citep[e.g.][]{1998A&A...337....9R,2006NewA...11..293A}.  We find that
the warp radius $r_{\rm w}$, the radius at which the disc bends away
from the inner flat disc is, in all cases, located well within the
optical radius, $R_{25}$, defined as the radius where $\mu_{\rm B} = 25$
mag/arcsec$^{2}$ \citep[c.f.][]{2010MNRAS.406..576P}. We find $r_{\rm
  w} \sim$ 9, 7, 8.5, 5, 10, 7 kpc for the five galaxies. In all
cases, however, we have $r_{w} \gtrsim R_{\rm d}$, where $R_{\rm d}$ is the
disc radial scale length (see Table~\ref{t1}). Warps of this type are
$\sim 35\%$ of of our simulated sample. Note that in all cases these
are S-shaped (`integral-sign') warps. Only for S6 we find a clearly
asymmetric warp. There is no clear example of a U-shaped warp in our
sample.

A third group of galaxies(specifically, S12, S13, S14, S15 and
S16, again $\sim 30 \%$ of our sample) show more complicated
vertical patterns reminiscent of that found by in Aq-C4 (see
G16).  In agreement with the empirically derived rules of
  \citet{1990ApJ...352...15B}, the patterns in many of these galaxies (e.g. Aq-C4, S14
  and S16) have a leading spiral morphology that winds with
gradually decreasing amplitude into the inner regions. In most cases,
these more complex vertical patterns result from the time evolution of
an initial integral-sign warp. After formation, such warps distort
into leading spirals as a result of the torque exerted by the inner,
misaligned disc \citep[e.g.][]{2006MNRAS.370....2S}. Note however that
a few discs present trailing spiral vertical patterns (e.g. S13 and
S15). We will explore the reasons for this in
Section~\ref{sec:hists}. The spiral morphology of these patterns is generally not as
regular and well defined as that of Aq-C4. Indeed, S13, S14 and
S16 could well be classified as S-shaped warps if they were to be
observed in certain edge-on orientations.  The vertical patterns in
this subset of galaxies resembles that observed in the outer MW disc
which, as previously described (see Section~\ref{sec:intro}), has an
oscillating asymmetry whose amplitude increases with Galactocentric
radius
\citep{2014ApJ...791....9S,2015MNRAS.452..676P,2015ApJ...801..105X}.

Thus, about $70\%$ of our simulated discs have clearly perturbed
vertical structures. The remaining $30\%$ are either weakly perturbed
or featureless.  This percentage agrees with observational
studies of the frequency of warps, although the definitions used in
these studies differ from and are not easily compared with those that
we use here \citep[e.g.][and references therein]{2006NewA...11..293A}.

\begin{figure}
    \includegraphics[width=40mm,clip]{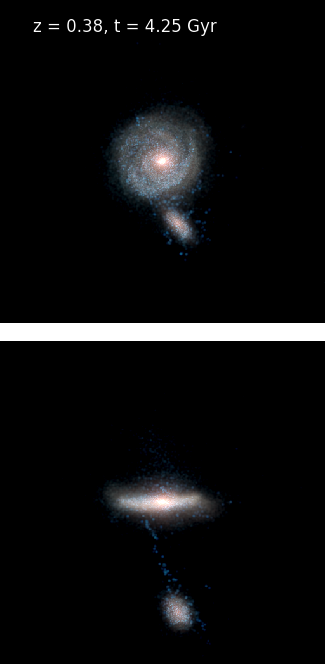}
    \includegraphics[width=40mm,clip]{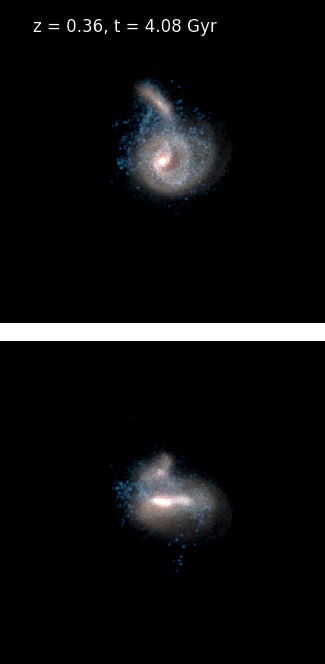}
    \caption{Face-on (top) and edge-on (bottom) projected stellar
      density for S3 at two different times, as indicated in the
      top left corner of the top panels. The images were constructed
      by mapping the K-, B- and U-band luminosities to the red, green
      and blue channels of a full colour composite image. Younger
      (older) star particles are therefore represented by bluer
      (redder) colours. Note that the outer regions of the
      pre-existing disc are severely perturbed during the head-on
      encounter with a satellite of $M_{\rm tot} = 10^{10.5}~\mo$.}
    \label{fig:Au12_snaps}
\end{figure}

\begin{figure}
    \centering
    \includegraphics[width=90mm,clip]{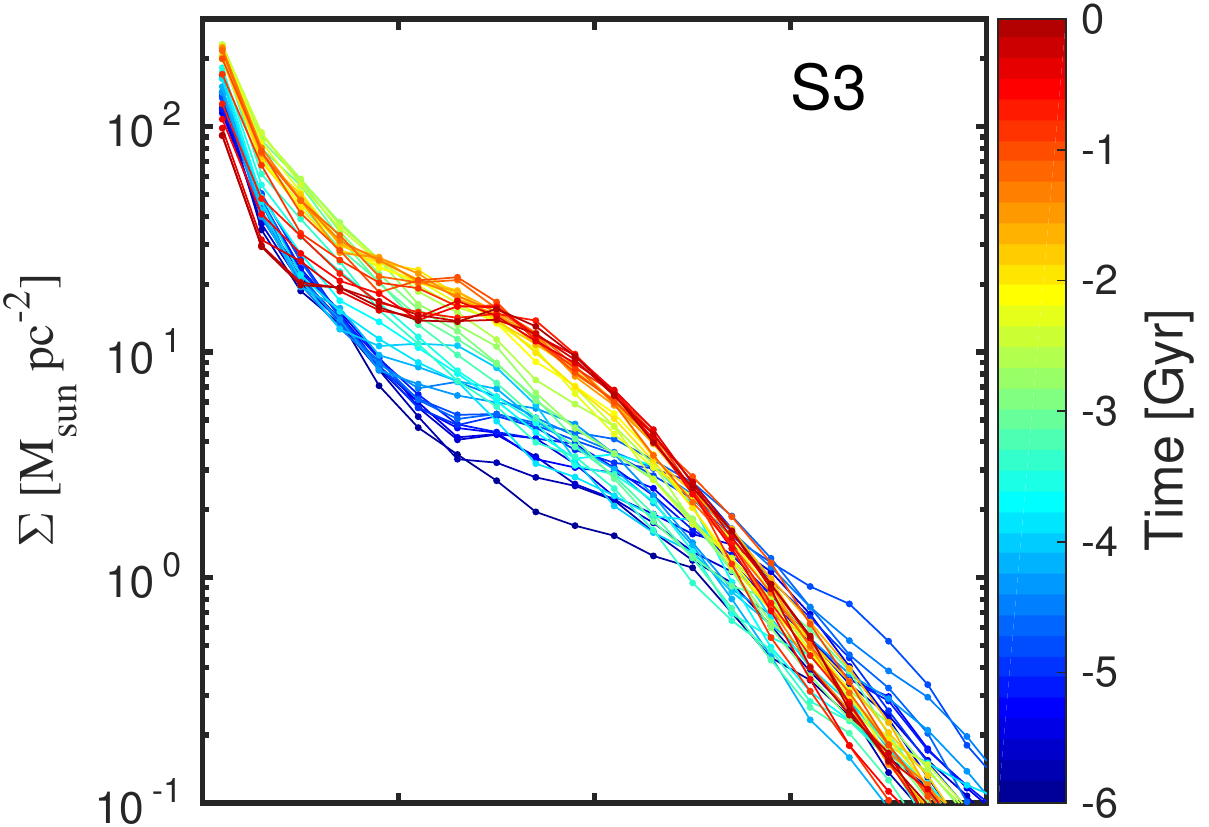}\\
    \vspace{0cm}
    \includegraphics[width=90mm,clip]{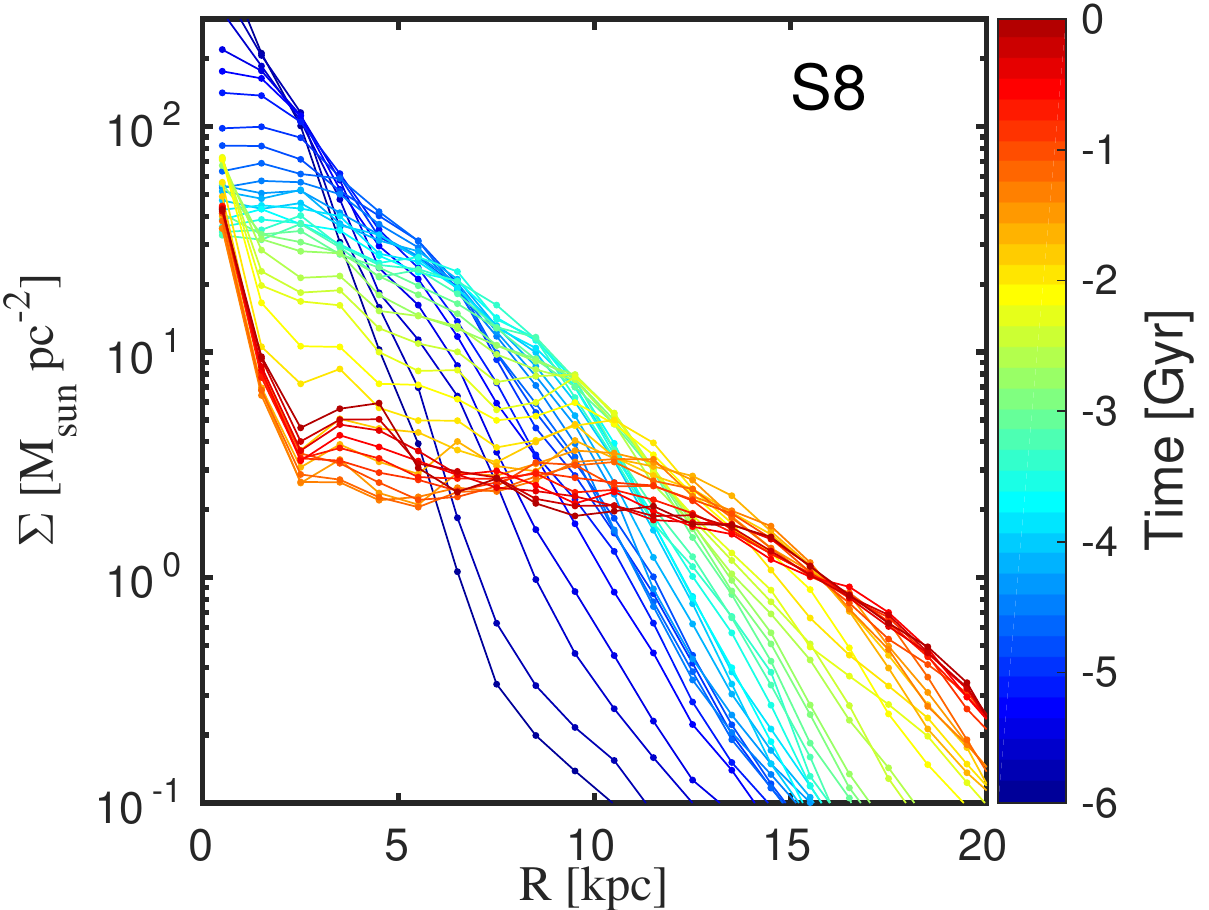}
    \caption{Radial profiles at different times of the mass surface density of star
      particles younger than 1 Gyr obtained from two different
      galaxies, as indicated on the top left corner of each
      panel. Only star particles within 2.5 kpc of the disc mid-plane
      are considered. The different colours indicate different
      times.}
    \label{fig:surf_dens}
\end{figure}

Figure~\ref{fig:maps_vz} shows mass-weighted mean vertical velocity
maps, \mvz, for each of our discs. These are equivalent to mean
(mass-weighted) line-of-sight velocity fields for perfectly face-on
galaxies. As expected, galaxies that show very small perturbations in
their \mz~maps also have relatively unperturbed \mvz~maps. However,
large perturbations can be seen in the remaining galaxies, with
$\Delta \langle {\rm V_{z}} \rangle = \langle {\rm V_{z}} \rangle_{\rm
  max} - \langle {\rm V_{z}} \rangle_{\rm min} > 60$ km/s in several
cases. Perturbations of this kind should be detectable in
line-of-sight stellar velocity fields of nearly face-on galaxies
obtained from integral field spectroscopy. For example, the Calar Alto
Legacy Integral Field Area Survey (CALIFA) provides such velocity
fields with uncertainties smaller than $\sim 20$ km/s
\citep{2014A&A...568A..70B}.  In addition, such vertical perturbations
should be visible in HI maps from surveys such as the Local Volume HI
Survey (LVHIS). The velocity fields of the galaxy HIPASS J1501-48,
presented in \citet{2015MNRAS.452.3139K}, clearly show that complex
spiral features can be observed in real face-on galaxies. The HI
line-of-sight velocity field of the galaxy UGC09965, recently
presented by \citet{2016A&A...585A..99M}, provides another
example. This galaxy, observed for the DiskMass survey
\citep{2010ApJ...716..198B}, also shows clear indications of a complex
vertical structure.

Comparison  of Figures~\ref{fig:maps_z} and  \ref{fig:maps_vz} reveals
the expected  anticorrelation between the absolute values  of \mz ~and
\mvz~ \citep[see][]{2013MNRAS.429..159G,G16}. This  is most evident in
warped galaxies, such  us S6 and S11, where we  see clearly that peaks
in $|$\mz$|$ correspond to regions  of $|$\mvz$| \sim 0$ km/s and vice
versa. More  wound  patterns, such  as  S14  and  S16 show  the  same
behavior.  Note  that in  the  approximation of  a tightly  wound
  bending wave $Z(R,\phi,t) =\Re[Z_{a}(R) {\rm e}^{im(\phi-\Omega_{\rm
      p}t)}]$,  where  $Z_{a}(R) =  Z(R)  {\rm  e}^{i\int{k dR}}$  and
  $\Omega_{\rm          p}$         the          pattern         speed
  \citep{2008gady.book.....B}.  Thus,  at  a  given  $R$,  $Z  \propto
  \Re[{\rm   e}^{im(\phi-\Omega_{\rm   p}t)}]$   and  $V_{z}   \propto
  \Re[im(\Omega   -  \Omega_{\rm   p})Z]$.  It   follows  that
line-of-sight  velocity  fields  of   face-on  discs,  such  as  those
presented here, can be used to reconstruct vertical structures even in
cases with more complicated \mz~morphologies such as S16.

It is interesting to explore how closely the vertical structure of the
cold gas traces the vertical structure of the stellar disc. Here, we
focus on the cold star-forming component identified by AREPO which
would correspond to the sum of the HI and H$_2$ components of real
galaxies. A detailed characterization of HI discs is presented in
a separate paper \citep{2016arXiv161001594M}.  In Figure~\ref{fig:maps_gz}
we show the \mz~maps obtained from this cold star-forming gas in each
galaxy.  Even though these maps reveal a significant amount of small
scale structure, in general, the gas follows the vertical patterns
seen in the stellar discs quite closely.  Discs with mildly perturbed
\mz~stellar maps, such as S2, S3 and S5, also show featureless
or weakly perturbed cold gas maps.  In contrast, systems with strongly
warped stellar discs (e.g. S6, S7 and S11) show very similar
patterns in the cold gas distribution.  Note that more complicated
vertical patterns, such as those in the stellar discs of S14 and
S16, are also reflected in the cold gas distributions.

Thus, we find, in most cases, that the vertical distortion patterns in
the stellar and gas discs are quite similar. This suggests either that
the patterns are tidally induced in all these cases \citep[see
  e.g.][]{1996AJ....111.1505C} or that differences in vertical
structure cannot be straightforwardly used to distinguish tidally
induced from accretion induced warps. We will explore this further in
Section~\ref{sec:hists}.

\section{Connecting present-day vertical structures with galactic histories} 
\label{sec:hists}

In this section we will connect the vertical patterns highlighted in
section~\ref{sec:pres-day} with the recent evolutionary histories of
individual galaxies.

As  previously   discussed,  only  $\sim  30\%$  of   our  discs  show
unperturbed  or weakly  perturbed  \mz~maps. In  the remaining  $70\%$
clear  vertical patterns  are seen  (see  Fig.~\ref{fig:maps_z}).  The
early  work   by  \citet{1969ApJ...155..747H}  showed   that  vertical
patterns, such as  those seen in our simulated discs,  can not be long
lived stable normal modes. Later attempts to model long lived vertical
modes based on the torque  exerted by a flattened and misaligned inner
dark  matter halo  \citep[e.g.][]{1988MNRAS.234..873S}  were shown  to
fail since  a misaligned inner  halo rapidly aligns with  the embedded
galactic disc \citep{1998MNRAS.297.1237B}.  As a result, such vertical
structures are likely to persist only  for a few Gyr and to be excited
by     external      perturbations     \cite[see][for     a     recent
review]{2013pss5.book..923S}.  Plausible perturbing agents are torques
exerted by  a misaligned outer dark matter  halo, misaligned accretion
of  cold gas,  direct tidal  perturbation by  orbiting  satellites, or
torques  exerted by  the dark  matter wakes  these  satellites produce
\citep[e.g.][]{1989MNRAS.237..785O,                1993ApJ...403...74Q,
  1998MNRAS.299..499W,    1999ApJ...513L.107D,    1999MNRAS.303L...7J,
  1999MNRAS.304..254V,    2000ApJ...534..598V,    2006MNRAS.370....2S,
  2010MNRAS.408..783R,2012MNRAS.426..983D,2014ApJ...789...90K,2014MNRAS.440.1971W,
  G16}.  In the following we try to identify the main perturbing agent
in each of our simulations.

\subsection{Mildly perturbed discs}

We start by  focusing on the subsample of galactic  discs that, at the
present-day, show relatively unperturbed vertical structure; i.e., S1,
S2,  S3,  S4 and  S5.  In  Figure~\ref{fig:tidal}  we show,  for  each
simulated  galaxy, the tidal  field exerted on the host by its 20 most massive satellites
as a function  of time, i.e. $|${\bf  a}$_{\rm sat}|  = GM_{\rm
  sat} / R_{\rm sat}^{3}$. Points are colour coded  according to the mass
of each satellite  at the corresponding time. This  figure shows that,
in general, the discs in  this subset have not interacted closely with
any satellite more massive that $M_{\rm min} \sim 10^{10}\mo$ over the
last 4 to  5 Gyr. It seems that large vertical  motions are not easily
excited  by companions less  massive than  $M_{\rm min}$.  This agrees
with  \citet{Grand16}, who finds  that a  disc heating  in at  least a
quarter  of these  simulations appears  associated with  satellites of
masses $  > M_{\rm min}$  but that smaller satellites  have negligible
effects.   Similar   results    have   been   recently   obtained   by
\citet{2015arXiv150803580M} and \citet{2015arXiv151101503D}.

S3  is an  interesting exception.  Unlike the  other galaxies  in this
subset,  it experiences  a violent  interaction $\sim  4$ Gyr  ago, an
almost  head-on encounter and  subsequent merger  with a  satellite of
mass $M_{\rm tot} \sim 10^{10.5}  \mo$. This nearly destroys the outer
regions of the disc, induces the formation of a strong bar 
\citep[see figure 5 of][]{Grand16} and significantly  heats  the surviving  disc. In  Figure
\ref{fig:Au12_snaps} we  can see  two simulation snapshots  during the
merger event.  As a result  of this interaction, significant inflow of
low  angular  momentum  cold  gas  into  the  inner  disc  regions  is
triggered,  substantially  boosting  the  star formation  rate  at  $R
\lesssim 10$ kpc \citep{1999IAUS..186..205M}.  This can be seen in the
top  panel  of  Figure~\ref{fig:surf_dens},  where we  show  the  time
evolution  of the  disc surface  density profile  obtained  from stars
that, at each time, are younger than 1 Gyr. By the present day, S3 has
a high  surface brightness,  compact disc, with  $R_{25} \sim  15$ kpc
(see  Figure~\ref{fig:maps_sb}).  Small  orbital  times in  this  disc
combine with a lack of recent perturbations to give a flat system with
no significant vertical asymmetries.

In Figure~\ref{fig:dm_orient} we follow the time evolution of the
alignment between each galactic disc and its surrounding dark
matter (DM) halo. As in G16, we compute, for every snapshot, the
DM halo inertia tensor,
\begin{equation}
\label{eqn:mten}
\mathcal{M}_{\alpha,\beta} = \frac{1}{M} \sum_{i=1}^{N_{p}} m_{i} r_{i,\alpha}r_{i,\beta},
\end{equation}
considering only the particles located within various spherical
shells. Here $N_{p}$ indicates the total number of particles within
the shell, $M$ their total mass, \textbf{\emph{r}}$_{i}$ and $m_{i}$
the position vector and mass of the {\it i}th particle, respectively.
$\alpha$ and $\beta$ are tensor indices taking values 1, 2 or 3.  We
consider four different spherical shells, defined by $0$~kpc~$ < r <
10$~kpc, $10$~kpc~$ < r < 25$~kpc, $25$~kpc~$ < r < 50$~kpc and
$50$~kpc$ < r < 80$~kpc.  We will refer to them as ${\rm dm}(a,b)$,
where $a$ and $b$ indicate the lower and upper radial limits of the
shell. Once computed, these tensors are diagonalized to obtain the
directions of the principal axes.

The different symbols in each panel of Figure~\ref{fig:dm_orient}
show, for each galaxy, the evolution of the angle between the total
disc angular momentum and the semi-minor axis of the various DM
shells.  In general, we find the two innermost DM shells, i.e. ${\rm
  dm}(0,10)$ and ${\rm dm}(10,25)$, to be well aligned with the disc
at all times, in good agreement with previous studies
\citep[e.g.][G16]{1998MNRAS.297.1237B, 2005ApJ...627L..17B,2012MNRAS.426..983D,2013MNRAS.428.1055A}.
At larger radii, some galaxies show increasing disc-halo
misalignment. Again, S3 is an extreme case in which the two outer
shells, ${\rm dm}(25,50)$ and ${\rm dm}(50,80)$, are severely
misaligned at all times shown. Note that the increasing misalignment
between the inner and outermost DM shells at $ -5 < t < -4$
Gyr correlates
well with the pericentric passage of the large perturbing satellite
previously discussed. Recall that this satellite penetrates deeply
into the host DM halo.  The two outermost layers continue to be
misaligned with the disc angular momentum until the present-day,
indicating that the torque exerted by these layers is not strong
enough to reorient this disc.  Similar correlations between features
in the misalignment between disc and outer halo and satellite
interactions can also be seen in other galaxies, for example, S1 at
$t \sim -2.5$ Gyr.

In Figure~\ref{fig:tilt_disk} we show the time evolution of the
orientation of each inner stellar disc (based on the total angular
momentum of disc particles with $r<5$~kpc\footnote{Our results are not
  sensitive to the exact definition here.}) with respect to its
orientation at $t = -5$ Gyr.  Excluding S3, all weakly
  perturbed or unperturbed discs show at most modest reorientation,
by angles $\lesssim 20^{\circ}$, over the last 5 Gyr \citep[see
  e.g.][]{2012MNRAS.426..983D,2014arXiv1411.3729Y}. S1 tilts by
$\sim 10^{\circ}$ during $-5 < t < -3.5$ Gyr, which correlates
well with the misalignment of the DM shell ${\rm dm}(10,25)$ during
this period.  Small jumps in these tilt angles are generally
correlated with interactions with intermediate mass satellites,
$10^{9.5} \lesssim M_{\rm sat} \lesssim 10^{10} \mo$, as in S1.
S3 shows one of the largest disc reorientations in the whole sample
which is due almost entirely to its merging with a satellite at
$t \approx -4$~Gyr with the concomitant near-destruction and
reformation of its disc, as already discussed in connection with
Figs~\ref{fig:Au12_snaps} and~\ref{fig:surf_dens}.  A secondary bump
can be seen during the last Gyr of evolution, which correlates with a
close encounter with a $\sim 10^{10}\mo$ satellite.

In summary, we find that, excluding S3, the galaxies with weak or
absent vertical structure have had a quiescent environment over the
last 5 Gyr. They show mild disc re-orientations and the inner shells
of their DM haloes are always well aligned with the galactic disc.  In
contrast, S3, suffered a violent interaction at $t \sim -4$
Gyr that nearly destroyed its outer disc and led to the formation of a newer
and compact inner disc which effectively resists subsequent
perturbations.

\subsection{Strongly perturbed discs}

\begin{figure*}
\centering
\includegraphics[width=180mm,clip]{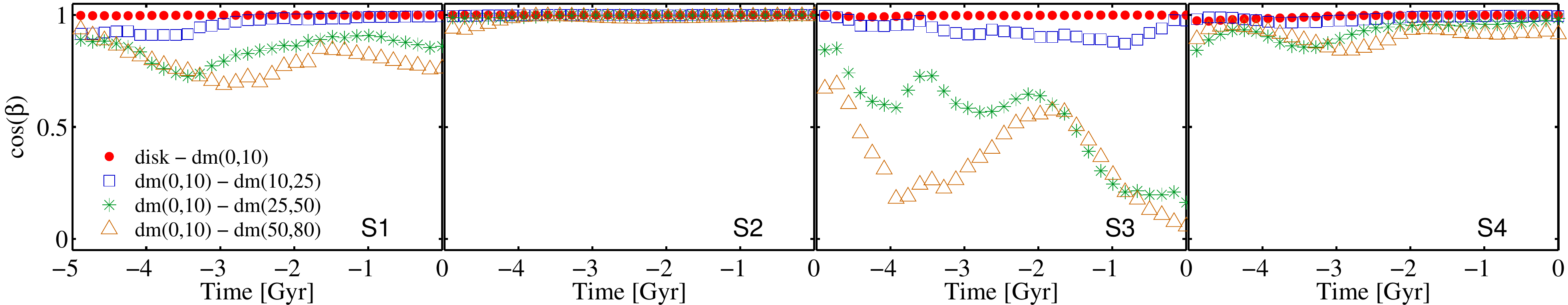}\\
\includegraphics[width=180mm,clip]{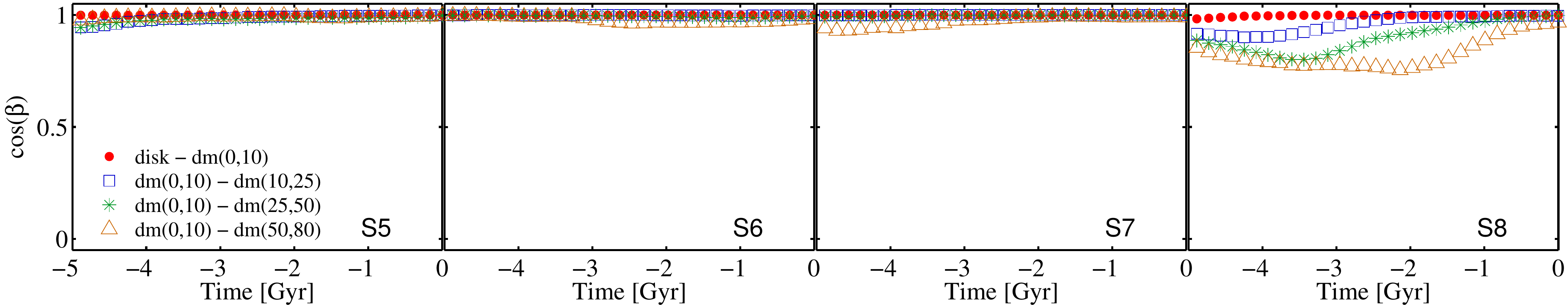}\\
\includegraphics[width=180mm,clip]{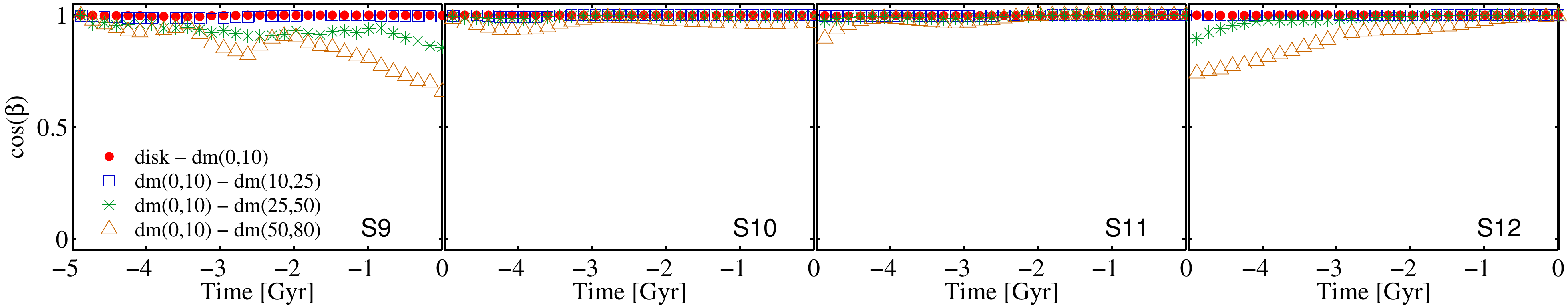}\\
\includegraphics[width=180mm,clip]{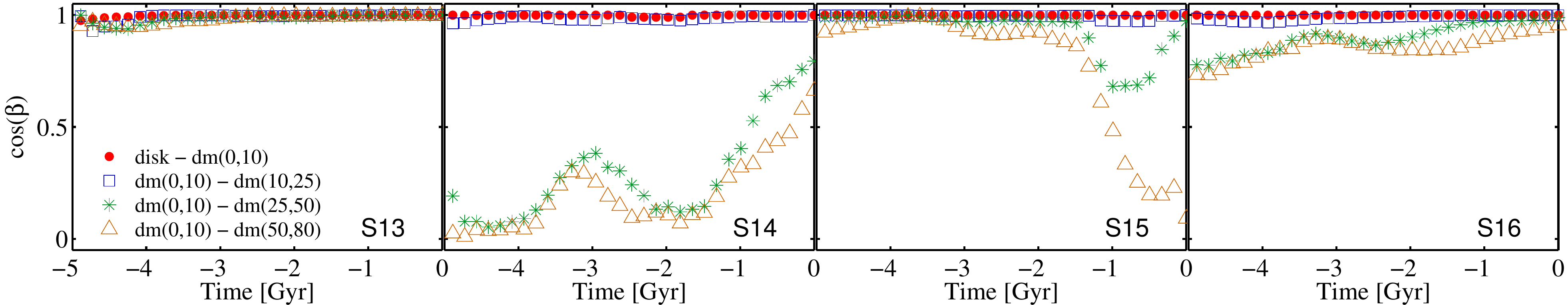}
\caption{The solid dots indicate the time evolution of the angle
  between the inner disc and the semi-minor axis of the inner DM halo
  (see Eq.~\ref{eqn:mten}), i.e. dm(0-10). Open symbols show the time
  evolution of the angle between the semi-minor axes of different DM
  shells, as indicated in the legend.}
\label{fig:dm_orient}
\end{figure*}

We now consider the discs that, at the present-day, show significantly
perturbed vertical structure. As discussed above, patterns in the
\mz~maps can be produced by a variety of mechanisms. Here we will not
attempt an exhaustive analysis of the history of perturbing torques on
a case by case basis, as done by G16 for the Aq-C4 simulation. Rather,
we will focus our analysis on general indicators that can help us to
identify the nature of the perturbers.

We start by exploring which of these galaxies have recently interacted
tidally with a massive companion. Figure~\ref{fig:tidal} shows the
time evolution of the tidal field exerted at system centre by each of
the 20 most massive satellites in each system. It demonstrates that
most of these galactic discs have had significant interactions with
satellites of $M_{\rm sat} \geq 10^{10.5}\mo$ during the last $\sim 4$
Gyr.  As shown in G16, even satellites with masses of $\sim
10^{10.5}~\mo$ and pericentric distances as large as 80 kpc can excite
strong vertical patterns in the disc either through their direct tidal
interaction or through an associated DM halo wake \citep[see
  also][]{2000ApJ...534..598V,2016MNRAS.457.2164O}.  Some of these
galaxies (in particular, S6, S7, S10, S11 and S15) have
interacted primarily with a single massive perturber during this
period.  The masses at infall of these perturbers are $11^{11.5},
10^{10.7}, 10^{11}, 10^{10.7}$ and $10^{11.5}~\mo$, respectively.  For
example, the massive perturber in S10 had its first pericentric
passage at $t = -6$ Gyr and undergoes a second close
interaction at $t = -4$ Gyr.  Other galaxies, such as S9,
S14 and S16, have been bombarded by two or more massive
satellites. The most massive perturber in these cases has an infall
mass of $10^{10.5}, 10^{11.1}$ and $10^{10.9}~\mo$, respectively.  On
the other hand S8, S12 and S13 show significantly perturbed
\mz~maps but no signs of close tidal interaction with perturbers of
$M_{\rm sat} \geq 10^{10.5}$. This suggests that other mechanisms 
play a role in these galaxies.

\begin{figure}
\centering
\includegraphics[width=85mm,clip]{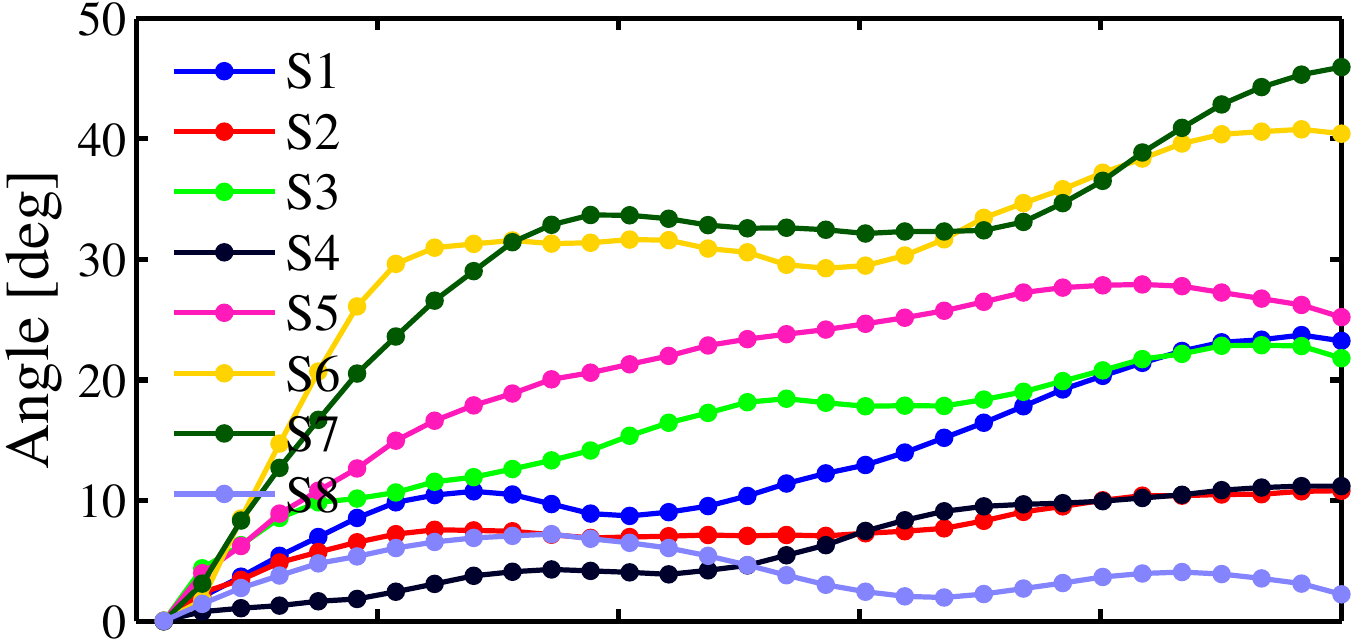}\\
\includegraphics[width=85mm,clip]{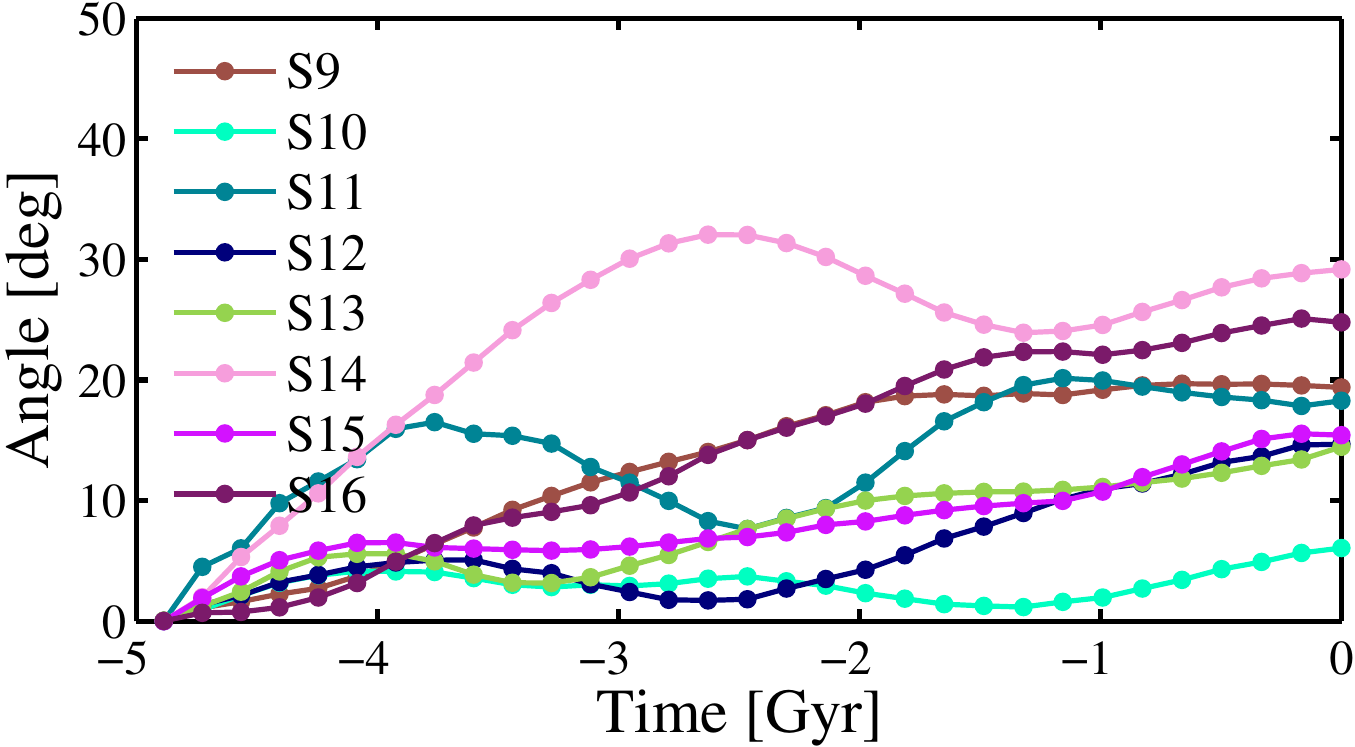}
\caption{Time evolution of the inner disc orientation with respect to
  its orientation at $t = -5$ Gyr.  Different colours
  correspond to different simulated galaxies, as indicated in the
  legend.}
\label{fig:tilt_disk}
\end{figure}

In Figure~\ref{fig:dm_orient} we explore whether misalignment between
different shells of DM could be driving any of these perturbations.
In general, galaxies that have interacted mainly with one massive
perturber show DM halos that are very well aligned with their galactic
disc, even up to distances as large as 80 kpc \citep[see
  also][]{2012MNRAS.426..983D}.  S6, S7, S10 and S11 show only
small bumps in the orientation of their outermost DM shells at the
time of maximum interaction, indicating that these satellites do not
trigger any long-term misalignment. However, we can see from
Figure~\ref{fig:tilt_disk} that these discs (and thus the systems as a
whole) do change orientation during these interactions. For example,
by the present-day, S6 shows a total disc tilt of $\sim 25^{\circ}$
since $t = -5$ Gyr.  The pericentric passage of the perturber
coincides with a dip in the time evolution of this tilting at $t \sim
-2.9$ Gyr.  S11 shows the largest tilt among this subset
of discs. Here also, the pericentric passage of the most significant
perturber coincides with a bump in the time evolution of the tilt (at
$t \sim -2.6$ Gyr). However, the disc shows ongoing
re-orientation well before this time, perhaps related to a strong
interaction with a different satellite that reached pericenter at
$t \sim -5.1$ Gyr.

\begin{figure}
    \includegraphics[width=40mm,clip]{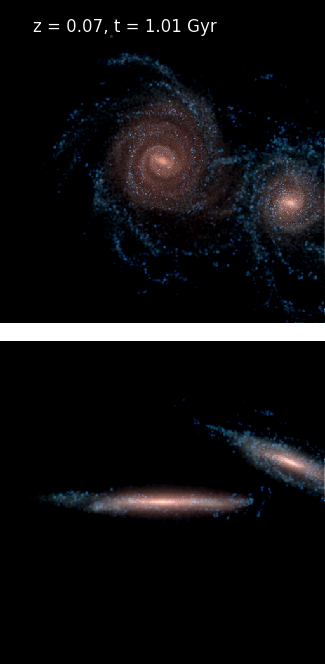}
    \includegraphics[width=40mm,clip]{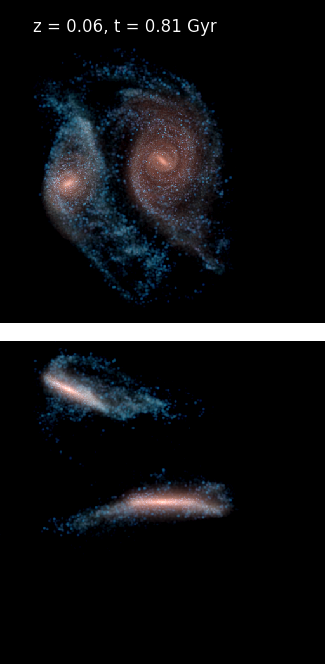}
    \caption{As in Figure~\ref{fig:Au12_snaps}, but for S15. At
      $t \sim -1$ Gyr this galaxy has a close encounter with a
      massive companion with an infall mass of $\sim 10^{11.5}~\mo$}
    \label{fig:Au25_snaps}
\end{figure}

In contrast, S15 develops a significant misalignment of its outer
halo shell (${\rm dm}(50,80)$) at late times.  As discussed earlier,
this galaxy undergoes a violent interaction with a $10^{11.5} \mo$
companion at $t \sim -0.9$ Gy; this correlates well with the
main reorientation of the DM shell.  Note that, unlike the
distant fly-by encounters experienced by S6, S7 and S11 (with
pericentric distances of $\sim 80, 70$ and 70 kpc respectively) the
S15 perturber penetrates deep into the host's inner 40 kpc \citep[see
Fig.14 of][]{Grand16}.  Figure~\ref{fig:Au25_snaps} shows two snapshots from
this strong interaction. The host disc
is tidally perturbed, inducing the formation of two tidal
arms. The vertical pattern seen in this simulation traces the trailing
structure of these tidal arms.

In general, galaxies that have suffered close encounters, (i.e. S3,
S9, S14, S15, and S16) tend to show more misaligned outer DM
shells. Features such as bumps and dips in the evolution of the
orientation of these shells correlate well with peak satellite
interactions.  Disc orientation variations also appear to correlate
with these interactions in a way which depends, among other things, on
the orbital properties of the satellite galaxies
\citep[e.g.,][]{1999MNRAS.304..254V, 2008MNRAS.391.1806V, 2009ApJ...700.1896K}.

While a tidal origin is clearly indicated for the vertical patterns in
many of our discs, other cases seem to require another explanation. A
specific example is S8, which shows a well-defined S-shaped warp
(see Fig.~\ref{fig:maps_z}). As can be seen from
Figure~\ref{fig:tidal}, this galaxy has not interacted closely with
any satellite for at least the last 5 Gyr. Nevertheless,
Figure~\ref{fig:dm_orient} reveals a significant misalignment of its
DM halo shells, especially at $t \lesssim -2$ Gyr. Even the
inner dm(10,25) DM shell is slightly misaligned with respect to the
disc plane. This misalignment correlates well with the tilting of the
disc over this period (Fig.~\ref{fig:tilt_disk}), suggesting that
torques exerted by the DM halo are driving the evolution. Over the
last 2 Gyr of evolution, however, the DM halo within 50 kpc aligns
well with the disc plane and the overall tilting of the system slows.
Despite this, we will see below that the warp in S8 grows
significantly over the final 0.5 Gyr. DM torques seem unlikely to be
the primary driver of this late evolution.

A mechanism which can excite vertical patterns in discs in the absence
of external tidal perturbations is the accretion of misaligned cold
gas \citep[see e.g.][]{2008A&ARv..15..189S,2013MNRAS.434.3142A}. For
example, \citet[][hereafter R10]{2010MNRAS.408..783R} analyse a
simulation in which a warp forms as a result of misalignment between
the inner stellar disc and a surrounding hot gas halo.  As discussed
in G16, in this situation, cold gas falling into the system is
strongly affected by the hot gas halo, and by the time it reaches the
disc, its angular momentum aligns with that of the hot gas, rather
than that of the disc. A misaligned outer disc then forms from the
newly accreted material. A characteristic of this mechanism is that it
does not produce significant warping of the pre-existing disc, and the
outer stellar warp is heavily dominated by newly formed stars. This
contrast with tidally induced warps which typically contain stars of
all ages (see e.g. G16).

This mechanism has recently been suggested for the formation of the
stellar warp in NGC 4565.  Using observation from the Hubble Space
Telescope, \citet{2014ApJ...780..105R} showed that stars older than 1
Gyr lie in a symmetrical distribution around the disc plane. In
contrast, a clear and strong warp can be seen in stellar populations
younger than 600 Myr and correlates well with the observed HI warp.

\begin{figure*}
\centering
\includegraphics[width=180mm,clip]{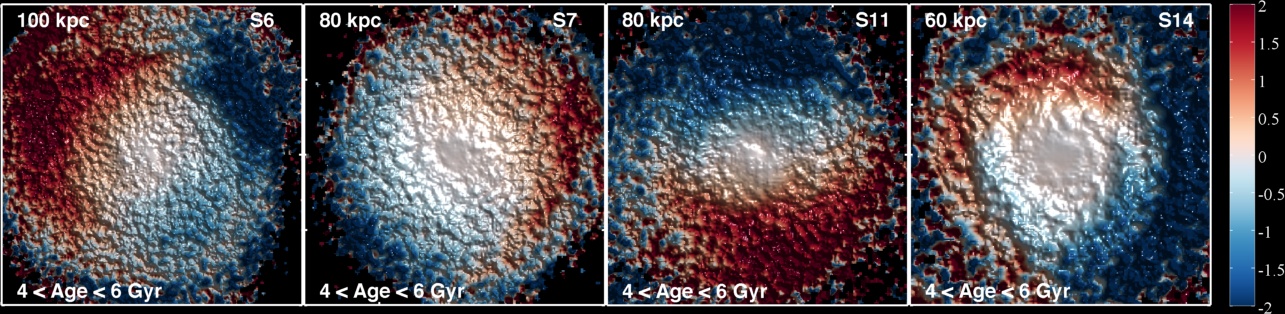}
\caption{Present-day maps of mass-weighted mean height, \mz, for four
  stellar discs, based purely on stars with ages between 4 and 6~Gyr.
  The colour-code and the relief show the value of \mz~in kpc as
  indicated by the colour bar. Each of the galaxies in this figure has
  undergone a strong tidal interactions with a massive satellite
  during the previous 4 Gyr.}
\label{fig:z_w}
\end{figure*}

In Figure~\ref{fig:z_w} we show \mz~maps obtained from stars with ages
between 4 and 6 Gyr in four discs that have undergone strong satellite
interactions during the last $\sim 4$ Gyr; namely S6, S7, S14
and S11. In all cases, very strong vertical patterns can be seen in
these older stellar populations. The shape and amplitude of these
patterns are very similar to those obtained when all stars are
included (see Fig.~\ref{fig:maps_z}), in good agreement with 
expectations for tidally excited features. Note that the
vertical patterns in these old populations also correlate well with those
seen in the distribution of cold star-forming gas. This
indicate that patterns in the gas and stellar components of a disc can
remain coincident for more than 1 Gyr. Similar results are found for
halos S9, S15 and S16.

\begin{figure*}
\centering
\includegraphics[width=180mm,clip]{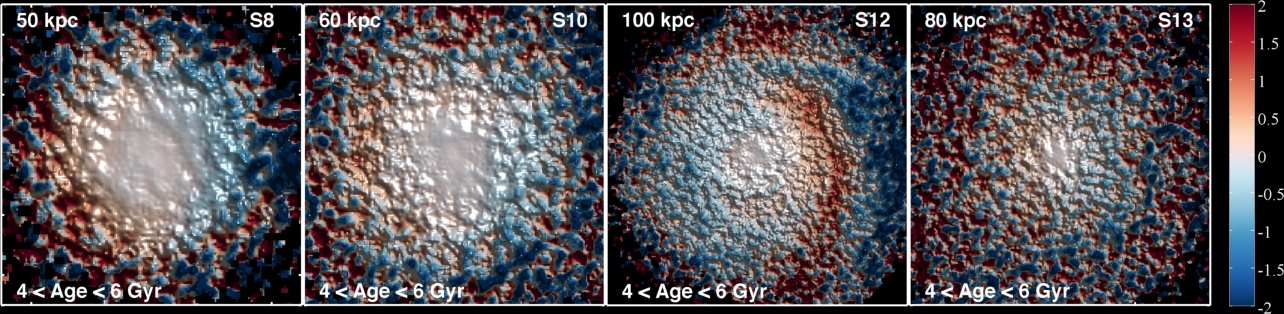}\\
\includegraphics[width=180mm,clip]{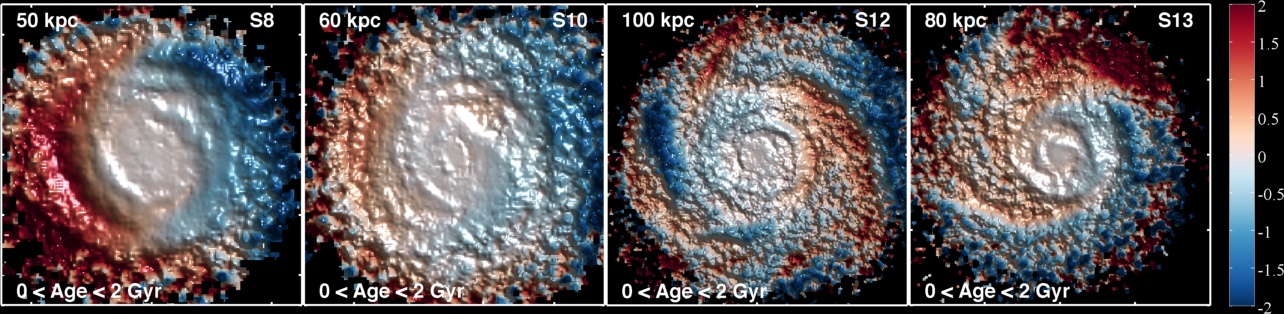}
\caption{Top panels: As in Figure~\ref{fig:z_w}, but for four galaxies
  that have not interacted strongly with a massive companions during
  the last 4 Gyr.  Bottom panels: Present-day \mz~maps for these same
  four galaxies, but now based on stars with ages between 0 and 2
  Gyr.}
\label{fig:z_nw}
\end{figure*}

\mz~maps for old star particles (ages between 4 and 6 Gyr) in the
remaining simulations with strong, present-day vertical patterns are
shown in the top panels of Figure~\ref{fig:z_nw}. With the exception
of S10, these galaxies have not interacted with any satellite more
massive than $10^{10.5}~\mo$ over the last $\sim 5$ Gyr. In contrast
to the galaxies just discussed, they show at most weak signatures of
vertical structure in their old stellar populations.  Some structure
is seen in S12, but it is much weaker and less complex than when all
stars are considered (compare with Fig.~\ref{fig:maps_z}). The bottom
panels of Figure~\ref{fig:z_nw} show \mz~maps for these same galaxies
constructed using only stars younger than 2 Gyr. These panels show
much stronger vertical patterns which correlate well with those found
for the full stellar populations (Fig.~\ref{fig:maps_z}) showing that
present-day vertical structure in these discs is dominated by young
stars.

S8 is a particularly clear example of a warp dominated by newly
formed stars. The bottom panel of Figure~\ref{fig:surf_dens} shows
surface density profiles for this disc made using star particles that,
at each time, are younger than 1 Gyr. Formation is evidently
inside-out, as expected for a disc that grows continuously by smooth
accretion of gas \citep[see, for example,][]{2006MNRAS.366..899N}. At
$t \sim -5$ Gyr, star formation is centrally concentrated.
However, as cold gas is consumed in the inner regions, a central
depression develops in the star formation rate (SFR). At later times,
stars are formed at progressively larger galactocentric distances from
newly accreted, high angular momentum gas, producing a dominant
population of young stars at large radii where the warp is seen. This
is the opposite of the behaviour discussed above for S3 (top panel
of Fig.~\ref{fig:surf_dens}) where a merger with a massive companion
at $t \sim -4$ Gyr triggered an inflow of cold gas to the
inner regions, boosting star formation there.

\begin{figure}
    \includegraphics[width=85mm,clip]{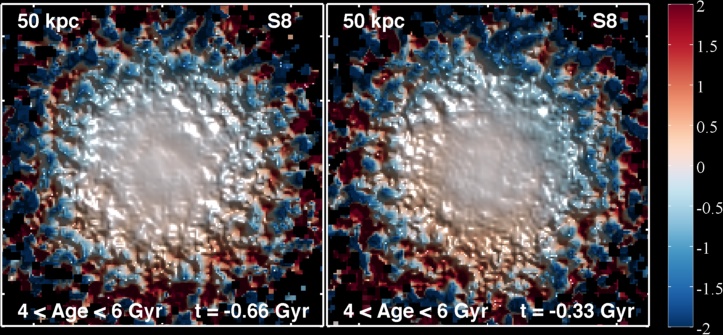}\\
    \includegraphics[width=85mm,clip]{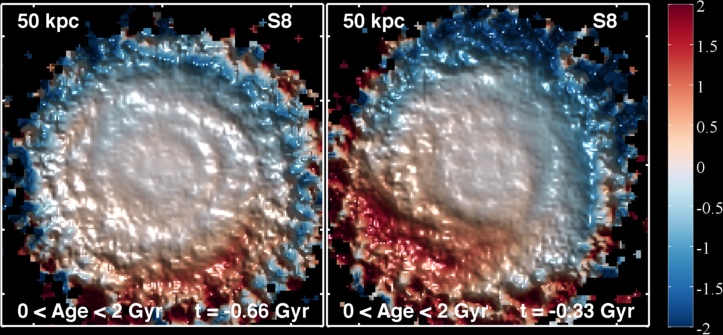}
    \caption{Maps of the mass-weighted mean height, \mz, for the S8
      disc at two different times, as indicated in the bottom right
      corner of each panel. The different colours and the relief
      indicate different values of \mz~in kpc. Top and bottom panels
      show the \mz~maps obtained from stars with ages between of 4 and
      6 Gyr, and between 0 and 2 Gyr, respectively.}
    \label{fig:H9_snaps}
\end{figure}

In Figure~\ref{fig:H9_snaps} we show \mz~maps of the S8 disc at two
recent times; $t = -0.33$ and $-0.66$ Gyr (left and right
columns, respectively). The bottom and top rows are based on stars
younger than 2 Gyr, and with ages between 4 and 6 Gyr, respectively.
While the warp in the younger component becomes stronger with time,
the older component remains quite flat.

\begin{figure}
    \includegraphics[width=40mm,clip]{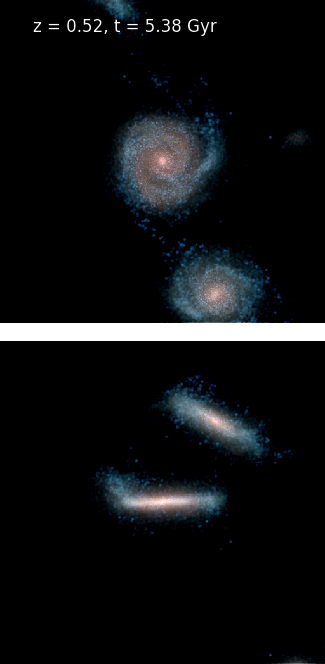}
    \includegraphics[width=40mm,clip]{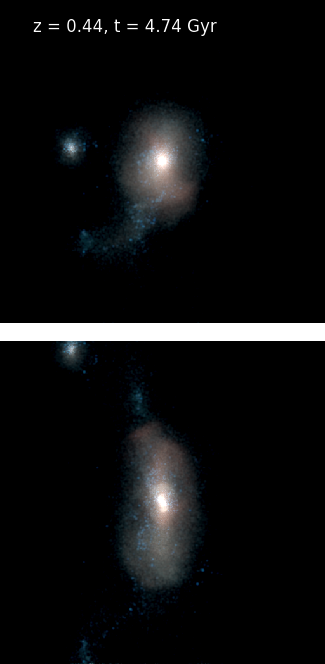}
    \caption{As in Figure~\ref{fig:Au12_snaps}, but for the galaxy
      S13. Note that after merging with a massive companion at
      $t \sim -5$ Gyr the pre-existing discs are completely
      destroyed.}
    \label{fig:Au19_snaps}
\end{figure}

S12 is another interesting case. Comparing Figures~\ref{fig:maps_z}
and \ref{fig:z_nw} we see that the \mz~map for the total stellar mass
combines the vertical patterns found in the older and the younger
populations, The former is apparently due to tidal effects, while the
latter results from accretion of misaligned gas. Note that the
\mz~maps found for the two subsets of stars are very different,

S13  and S10  are  rather similar  to  S3. Each  merges with  a
massive companion (at $t = -5$ and 4.2 Gyr, respectively). The disc of
S10 (which has  $R_{25} \sim 17$ kpc at  $t \sim -5$ Gyr)
is strongly perturbed by the  merger and forms a strong bar.  However,
unlike S3,  it continues to  accrete high angular momentum  gas and
regrows  an extended  disc by  the  present-day.  This  is visible  in
Figure  2 of  \citet{Grand16},  where stellar  surface  density is  plotted as  a
function of time.  The warps  in the young stellar populations of this
galaxy are  again due to  misaligned gas accretion.  In  contrast, the
pre-existing  disc of S13  is completely  destroyed by  its gas-rich
merger,  but a  new  thin disc  quickly  forms thereafter  due to  the
re-accretion          of          misaligned         gas          (see
Figure~\ref{fig:Au19_snaps}). Unlike  S8, where the  accreted gas is
new  material arriving  from large  radius, the  newly formed  disc in
S13 is made from gas previously associated with the merging galaxies
\citep{1992AJ....104.1039S, 2005ApJ...622L...9S}. This misaligned accreted  gas, which
  shares the sense of  rotation of the newly formed disc\footnote{ See
    \citet{2013MNRAS.434.3142A} for the formation of counter-rotating
    discs on top of  pre-existing discs due to misaligned infall of gas.}, develops
  a vertical pattern with a trailing spiral morphology.

In summary, we find that vertical structure in our discs is generated
by a variety of mechanisms. Close encounters with large companions
result in tidally induced patterns that are present in both young and
old stellar populations.  Mergers with massive and gas-rich satellites
can severely perturb or even destroy pre-existing discs, but
subsequent accretion of misaligned gas can produce new discs with
vertical features that are significantly stronger in the youngest
stellar populations. We have also found warps in galaxies that have
not interacted with any massive satellite for at least 5 Gyr.  Such
patterns are generally the result of misaligned gas accretion either
from new infall or from residual gas from an earlier gas-rich merger.

\section{Summary and Conclusions}
\label{sec:summ}

In this study we have characterized the vertical structure of galactic
stellar  discs in  a suite  of fully  cosmological simulations  of the
formation and evolution of  individual Milky Way-like galaxies.  Based
on  the state-of-the-art,  $N$-body magnetohydrodynamical  code AREPO,
the  galactic  discs  in  these  simulations  show  a  wide  range  in
morphology  and extent  at  $z=0$, including  strong bars,  flocculent
multi-arm  spirals, grand  design  spirals, and  compact high  surface
brightness discs.  The simulated Milky Way sized systems were selected
to be  located at  least 1.37 Mpc  from any  object more than  half of
their mass. In  addition, we have focused our analysis  on a subset of
haloes that exhibit  clear stellar discs at the  present-day, in order
to characterize their vertical structure.

For each galactic disc, we  have constructed maps of the mass-weighted
mean   height,  \mz,   for  both   the  stellar   and  the   cold  gas
components. Approximately $30\%$ show little or no vertical structure,
but the  remainder show clear vertical patterns,  with amplitudes that
can be as large as  2 kpc.  This fraction of $70\%$\footnote{Including
  the  simulation Aq-C4  studied in  G16.}  agrees  with observational
studies      of     large      samples     of      edge-on     spirals
\citep[e.g.][]{2006NewA...11..293A}.   We   identify  four  mechanisms
producing such  structure: close satellite  encounters, distant flybys
of  massive   companions,  accretion  of  misaligned   cold  gas,  and
re-accretion  of cold  gas from  the progenitors  of a  gas-rich major
merger.  Half  of our vertically structured discs  show ``S-shape'' or
``integral sign'' warps, with  a distribution of warp angle, $\alpha$,
centered around  $\langle \alpha  \rangle = 4^{\circ}  \pm 2^{\circ}$,
which      agrees      well      with     observational      estimates
\citep[e.g.][]{1998A&A...337....9R,2006NewA...11..293A}. In all cases,
we find  the warp to begin  well within $R_{25}$ but  outside one disc
scale-length  \citep[c.f.][]{2010MNRAS.406..576P}. There  is  no clear
example  of a  U-shaped  warp  in our  sample.   The other  vertically
structured  discs show  more complex  patterns, mostly  with  a spiral
morphology that  winds into the inner  disc and is  reminiscent of the
structure         seen         in         the        Milky         Way
\citep{2014ApJ...791....9S,2015MNRAS.452..676P,2015ApJ...801..105X,2016arXiv160407501M}. Including
the simulation Aq-C4  from G16, $35\%$ of our  simulated discs show an
oscillating vertical asymmetry,  suggesting that such behaviour should
be quite common.   A resolution study performed on  two galactic discs
shows that the amplitude  and morphology of the vertical perturbations
are initially well converged.  Nonetheless, the damping time-scales of
the patterns  can differ  at different resolution  levels. It  is also
important  to highglight  that, overall,  our final  simulated stellar
discs   are  thicker   than   observed  \citep[$h_{\rm   z}  \sim   1$
kpc\footnote{Note  that the  young stellar  discs show  a  much thiner
  vertical  distribution},  see][]{2016arXiv161001159G}.  Thus,  their
vertical rigidity  is significantly lower  than that of the  Milky Way
disc.

We have also explored how closely vertical structure in the cold star
forming gas traces that in disc stars. In general, we find that gas
and stars follow the same overall pattern, albeit with significant
differences on small scales. This is true even for discs with complex
oscillatory patterns, and appears independent of the mechanism driving
the vertical perturbations. Our results also indicate that the
vertical structure of the cold gas and the stellar discs can remain
coincident for more than 1 Gyr.

Maps of the stellar mass-weighted mean vertical velocity \mvz~confirm the
oscillatory behavior of these features. Galaxies with strongly
perturbed \mz~maps also have significantly perturbed \mvz~maps, and a
clear anticorrelation between the absolute values of \mz~and
\mvz~is visible in
all cases \citep[see also][]{2013MNRAS.429..159G,G16}.  Thus,
\mvz~maps can be used to reconstruct vertical structure in a disc,
even for complicated morphologies. The vertical velocity perturbations
can be as large as $\Delta \langle {\rm V_{z}} \rangle = \langle {\rm
  V_{z}} \rangle_{\rm max} - \langle {\rm V_{z}} \rangle_{\rm min} >
60$ km/s.  Such perturbations should be easily detectable in nearly
face-on galaxies either from line-of-sight stellar velocity fields
obtained by integral field spectroscopy, or from cold gas velocity
fields obtained by radio interferometry. Such observations thus
provide a direct way to assess the frequency with which oscillating
vertical patterns arise in real late-type galaxies.

We have explored the connection between recent assembly history and
present-day vertical structure for our suite of Milky Way-like
simulations. We find that most (but not all) discs with little or no
vertical structure have had no interaction with a satellite more
massive than $M_{\rm min} \sim 10^{10}~\mo$ over the last 4 to 5
Gyr. It appears that strong, long-lasting vertical patterns are
rarely excited by satellites of mass $\lesssim M_{\rm min}$. Similar
results based on idealized but cosmologically motivated simulations
have recently been reported by \citet{2015arXiv151101503D}. This is
also in agreement with \citet{Grand16} who finds that such
perturbations are a significant source of disc heating in at least a
quarter of our simulations, and that satellites less massive than
$M_{\rm min}$ play a negligible role in disc heating \citep[see
  also][]{2015arXiv150803580M}. Note, however, that $M_{\rm min}$
should not be interpreted as a rigid mass threshold; whether or not a
large-scale bending pattern is excited depends not only on the perturber
mass but also on its pericentric distance and velocity
\citep{2014MNRAS.440.1971W,2015arXiv151101503D}.

We note that a present-day vertically relaxed disc does not
necessarily imply a quiescent evolutionary history over the last $\sim
5$ Gyr. In particular, our sample includes one example (S3) of a
galaxy that merged with a $10^{10.5}~\mo$ satellite $\sim 4$ Gyr
ago. This merger destroyed the outer regions of the pre-existing disc,
and triggered the inflow of residual low angular momentum gas to the
inner disc, substantially boosting the star formation rate within 10
kpc. As a result, the present-day disc is compact and has high surface
density, allowing it to come to equilibrium well before $z=0$.
 
Most galaxies with strong, present-day vertical patterns have
interacted at least once with a satellite more massive than
$10^{10.5}~\mo$ during the last 4 to 5 Gyr. As shown by previous
studies, low-velocity and relatively distant encounters with
satellites with masses as small as $10^{10.5}~\mo$ can significantly
perturb a galactic disc thanks both to direct tidal interaction and to
the strong response such flybys can excite in the galaxy's DM halo
\citep[e.g.][]{2000ApJ...534..598V,G16}.

We have examined whether a misalignment between discs and their DM
haloes could be the main driver behind any of the warps in our
simulations.  In agreement with previous studies
\citep[e.g.][]{1998MNRAS.297.1237B, 2005ApJ...627L..17B,2012MNRAS.426..983D,2013MNRAS.434.3142A,G16},
we find that, in almost all cases, disc angular momentum is well
aligned with the semi-minor axis of the dark matter halos within their
inner 25 kpc, independent of the present-day vertical structure.
Interestingly, galaxies that have experienced massive flybys often
show DM halos that are well aligned out to distances as large as 80
kpc (S6, S7, S11).  This suggests that disc-halo misalignment is
not the main driver of their vertical perturbations.  In such cases,
we find that the satellites do not penetrate far into the host halo.
In contrast, galaxies that undergo close encounters with massive
companions tend to show significantly misaligned outer DM shells
(e.g. S3, S14, S15, S16). Features such as bumps and dips in
the evolving orientation of these shells are well correlated with the
encounters. Note, however, that although the misaligned outer halo
provides a torque on the disc in addition to that from the satellite,
the effect due to regions beyond $\sim 50$~kpc is negligible in
comparison to that from the satellite and the inner halo (see
e.g. G16).

A few of our simulations show prominent vertical structures without
any recent interaction with a satellite of mass $> M_{\rm min}$
(e.g. S8 and S13). Other mechanisms, in particular, the accretion
of misaligned cold gas
\citep[e.g.][]{2008A&ARv..15..189S,2013MNRAS.428.1055A,2010MNRAS.408..783R},
are playing a role in these systems. We show that while the vertical
patterns are present both in young stars (ages less than 2~Gyr) and
in the cold gas in these two galaxies, they are almost absent in
the older stars (ages between 4 and 6 Gyr). This contrasts with
tidally induced structures, which (in other systems) are almost
equally present in stars of all ages.\footnote{Note however that
  vertical patterns tend to wash out somewhat faster in kinematically
  hotter populations (G16).}  This is as expected for misaligned
accretion either of newly accreted halo gas (e.g. S8) or of cold gas
left over after a gas-rich merger (e.g. S13).  In these situations
new stars form from the (re)accreted gas and older stars are at most
weakly affected.

Several studies have discussed plausible mechanisms to explain the
North-South asymmetry observed in the Milky Way's disc; in particular,
the Monoceros ring.  Some of these papers suggest that this structure
could be the result of the tidal disruption of a satellite galaxy on
an almost coplanar and nearly circular orbit
\citep[e.g.][]{2003ApJ...592L..25H,2005ApJ...626..128P,2011MNRAS.414L...1M}. Interestingly,
none of the vertical oscillatory patterns seen in our set of
simulations is associated with a tidally disrupted satellite
galaxy. Rather, the vertical asymmetries are traced by stars formed
in-situ in all cases. Thus, our study suggests that tidal disruption
of this kind is less likely as an origin for the North-South asymmetry
of the Galactic disc than the other mechanisms we have considered.

\section*{Acknowledgements}

We  would  like to  thank  the referee,  James  Binney, for  an
insightful report  that led to  improvements in the  presentation and
content of this  paper. RG and VS acknowledge  support through the DFG
Research Centre SFB-881 'The Milky Way System' through project A1.  VS
and  RP acknowledges support  by the  European Research  Council under
ERC-StG grant EXAGAL-308037.

%%%%%%%%%%%%%%%%%%%%%%%%%%%%%%%%%%%%%%%%%%%%%%%%%%

%%%%%%%%%%%%%%%%%%%% REFERENCES %%%%%%%%%%%%%%%%%%

\bibliographystyle{mn2e}
\bibliography{stat_disks}

\begin{thebibliography}{76}
\expandafter\ifx\csname natexlab\endcsname\relax\def\natexlab#1{#1}\fi

\bibitem[{{Ann} \& {Park}(2006)}]{2006NewA...11..293A}
{Ann} H.~B., {Park} J.-C., 2006, \na, 11, 293

\bibitem[{{Aumer} \& {White}(2013)}]{2013MNRAS.428.1055A}
{Aumer} M., {White} S.~D.~M., 2013, \mnras, 428, 1055

\bibitem[{{Aumer} {et~al}\mbox{.}(2013){Aumer}, {White}, {Naab}, \&
  {Scannapieco}}]{2013MNRAS.434.3142A}
{Aumer} M., {White} S.~D.~M., {Naab} T., {Scannapieco} C., 2013, \mnras, 434,
  3142

\bibitem[{{Bailin}(2003)}]{2003ApJ...583L..79B}
{Bailin} J., 2003, \apjl, 583, L79

\bibitem[{{Bailin} {et~al}\mbox{.}(2005){Bailin}, {Kawata}, {Gibson},
  {Steinmetz}, {Navarro}, {Brook}, {Gill}, {Ibata}, {Knebe}, {Lewis}, \&
  {Okamoto}}]{2005ApJ...627L..17B}
{Bailin} J. {et~al.}, 2005, \apjl, 627, L17

\bibitem[{{Barrera-Ballesteros} {et~al}\mbox{.}(2014){Barrera-Ballesteros},
  {Falc{\'o}n-Barroso}, {Garc{\'{\i}}a-Lorenzo}, {van de Ven}, {Aguerri},
  {Mendez-Abreu}, {Spekkens}, {Lyubenova}, {S{\'a}nchez}, {Husemann}, {Mast},
  {Garc{\'{\i}}a-Benito}, {Iglesias-Paramo}, {Del Olmo}, {M{\'a}rquez},
  {Masegosa}, {Kehrig}, {Marino}, {Verdes-Montenegro}, {Ziegler}, {McIntosh},
  {Bland-Hawthorn}, {Walcher}, \& {Califa Collaboration}}]{2014A&A...568A..70B}
{Barrera-Ballesteros} J.~K. {et~al.}, 2014, \aap, 568, A70

\bibitem[{{Bershady} {et~al}\mbox{.}(2010){Bershady}, {Verheijen}, {Swaters},
  {Andersen}, {Westfall}, \& {Martinsson}}]{2010ApJ...716..198B}
{Bershady} M.~A., {Verheijen} M.~A.~W., {Swaters} R.~A., {Andersen} D.~R.,
  {Westfall} K.~B., {Martinsson} T., 2010, \apj, 716, 198

\bibitem[{{Binney}(1992)}]{1992ARA&A..30...51B}
{Binney} J., 1992, \araa, 30, 51

\bibitem[{{Binney}, {Jiang} \& {Dutta}(1998){Binney}, {Jiang}, \&
  {Dutta}}]{1998MNRAS.297.1237B}
{Binney} J., {Jiang} I.-G., {Dutta} S., 1998, \mnras, 297, 1237

\bibitem[{{Binney} \& {Tremaine}(2008)}]{2008gady.book.....B}
{Binney} J., {Tremaine} S., 2008, {Galactic Dynamics: Second Edition}.
  Princeton University Press

\bibitem[{{Briggs}(1990)}]{1990ApJ...352...15B}
{Briggs} F.~H., 1990, \apj, 352, 15

\bibitem[{{Cox} {et~al}\mbox{.}(1996){Cox}, {Sparke}, {van Moorsel}, \&
  {Shaw}}]{1996AJ....111.1505C}
{Cox} A.~L., {Sparke} L.~S., {van Moorsel} G., {Shaw} M., 1996, \aj, 111, 1505

\bibitem[{{de la Vega} {et~al}\mbox{.}(2015){de la Vega}, {Quillen}, {Carlin},
  {Chakrabarti}, \& {D'Onghia}}]{2015MNRAS.454..933D}
{de la Vega} A., {Quillen} A.~C., {Carlin} J.~L., {Chakrabarti} S., {D'Onghia}
  E., 2015, \mnras, 454, 933

\bibitem[{{Debattista} \& {Sellwood}(1999)}]{1999ApJ...513L.107D}
{Debattista} V.~P., {Sellwood} J.~A., 1999, \apjl, 513, L107

\bibitem[{{DeBuhr}, {Ma} \& {White}(2012){DeBuhr}, {Ma}, \&
  {White}}]{2012MNRAS.426..983D}
{DeBuhr} J., {Ma} C.-P., {White} S.~D.~M., 2012, \mnras, 426, 983

\bibitem[{{D'Onghia} {et~al}\mbox{.}(2015){D'Onghia}, {Madau}, {Vera-Ciro},
  {Quillen}, \& {Hernquist}}]{2015arXiv151101503D}
{D'Onghia} E., {Madau} P., {Vera-Ciro} C., {Quillen} A., {Hernquist} L., 2015,
  ArXiv e-prints

\bibitem[{{G{\'o}mez} {et~al}\mbox{.}(2013){G{\'o}mez}, {Minchev}, {O'Shea},
  {Beers}, {Bullock}, \& {Purcell}}]{2013MNRAS.429..159G}
{G{\'o}mez} F.~A., {Minchev} I., {O'Shea} B.~W., {Beers} T.~C., {Bullock}
  J.~S., {Purcell} C.~W., 2013, \mnras, 429, 159

\bibitem[{{G{\'o}mez} {et~al}\mbox{.}(2016){G{\'o}mez}, {White}, {Marinacci},
  {Slater}, {Grand}, {Springel}, \& {Pakmor}}]{G16}
{G{\'o}mez} F.~A., {White} S.~D.~M., {Marinacci} F., {Slater} C.~T., {Grand}
  R.~J.~J., {Springel} V., {Pakmor} R., 2016, \mnras, 456, 2779

\bibitem[{{Grand} {et~al}\mbox{.}(2016{\natexlab{a}}){Grand}, {G{\'o}mez},
  {Marinacci}, {Pakmor}, {Springel}, {Campbell}, {Frenk}, {Jenkins}, \&
  {White}}]{2016arXiv161001159G}
{Grand} R.~J.~J. {et~al.}, 2016{\natexlab{a}}, ArXiv e-prints

\bibitem[{{Grand} {et~al}\mbox{.}(2016{\natexlab{b}}){Grand}, {Springel},
  {G{\'o}mez}, {Marinacci}, {Pakmor}, {Campbell}, \& {Jenkins}}]{Grand16}
{Grand} R.~J.~J., {Springel} V., {G{\'o}mez} F.~A., {Marinacci} F., {Pakmor}
  R., {Campbell} D.~J.~R., {Jenkins} A., 2016{\natexlab{b}}, \mnras, 459, 199

\bibitem[{{Helmi} {et~al}\mbox{.}(2003){Helmi}, {Navarro}, {Meza}, {Steinmetz},
  \& {Eke}}]{2003ApJ...592L..25H}
{Helmi} A., {Navarro} J.~F., {Meza} A., {Steinmetz} M., {Eke} V.~R., 2003,
  \apjl, 592, L25

\bibitem[{{Hunter} \& {Toomre}(1969)}]{1969ApJ...155..747H}
{Hunter} C., {Toomre} A., 1969, \apj, 155, 747

\bibitem[{{Jiang} \& {Binney}(1999)}]{1999MNRAS.303L...7J}
{Jiang} I.-G., {Binney} J., 1999, \mnras, 303, L7

\bibitem[{{Kaiser} {et~al}\mbox{.}(2010){Kaiser}, {Burgett}, {Chambers},
  {Denneau}, {Heasley}, {Jedicke}, {Magnier}, {Morgan}, {Onaka}, \&
  {Tonry}}]{2010SPIE.7733E..0EK}
{Kaiser} N. {et~al.}, 2010, in Society of Photo-Optical Instrumentation
  Engineers (SPIE) Conference Series, Vol. 7733, Society of Photo-Optical
  Instrumentation Engineers (SPIE) Conference Series, p.~0

\bibitem[{{Kamphuis} {et~al}\mbox{.}(2015){Kamphuis}, {J{\'o}zsa}, {Oh},
  {Spekkens}, {Urbancic}, {Serra}, {Koribalski}, \&
  {Dettmar}}]{2015MNRAS.452.3139K}
{Kamphuis} P., {J{\'o}zsa} G.~I.~G., {Oh} S.-.~H., {Spekkens} K., {Urbancic}
  N., {Serra} P., {Koribalski} B.~S., {Dettmar} R.-J., 2015, \mnras, 452, 3139

\bibitem[{{Kazantzidis} {et~al}\mbox{.}(2008){Kazantzidis}, {Bullock},
  {Zentner}, {Kravtsov}, \& {Moustakas}}]{2008ApJ...688..254K}
{Kazantzidis} S., {Bullock} J.~S., {Zentner} A.~R., {Kravtsov} A.~V.,
  {Moustakas} L.~A., 2008, \apj, 688, 254

\bibitem[{{Kazantzidis} {et~al}\mbox{.}(2009){Kazantzidis}, {Zentner},
  {Kravtsov}, {Bullock}, \& {Debattista}}]{2009ApJ...700.1896K}
{Kazantzidis} S., {Zentner} A.~R., {Kravtsov} A.~V., {Bullock} J.~S.,
  {Debattista} V.~P., 2009, \apj, 700, 1896

\bibitem[{{Kim} {et~al}\mbox{.}(2014){Kim}, {Peirani}, {Kim}, {Ann}, {An}, \&
  {Yoon}}]{2014ApJ...789...90K}
{Kim} J.~H., {Peirani} S., {Kim} S., {Ann} H.~B., {An} S.-H., {Yoon} S.-J.,
  2014, \apj, 789, 90

\bibitem[{{L{\'o}pez-Corredoira} {et~al}\mbox{.}(2002){L{\'o}pez-Corredoira},
  {Cabrera-Lavers}, {Garz{\'o}n}, \& {Hammersley}}]{2002A&A...394..883L}
{L{\'o}pez-Corredoira} M., {Cabrera-Lavers} A., {Garz{\'o}n} F., {Hammersley}
  P.~L., 2002, \aap, 394, 883

\bibitem[{{Marinacci} {et~al}\mbox{.}(2016){Marinacci}, {Grand}, {Pakmor},
  {Springel}, {G{\'o}mez}, {Frenk}, \& {White}}]{2016arXiv161001594M}
{Marinacci} F., {Grand} R., {Pakmor} R., {Springel} V., {G{\'o}mez} F., {Frenk}
  C., {White} S., 2016, ArXiv e-prints

\bibitem[{{Marinacci}, {Pakmor} \& {Springel}(2014){Marinacci}, {Pakmor}, \&
  {Springel}}]{2014MNRAS.437.1750M}
{Marinacci} F., {Pakmor} R., {Springel} V., 2014, \mnras, 437, 1750

\bibitem[{{Marinacci} {et~al}\mbox{.}(2015){Marinacci}, {Vogelsberger}, {Mocz},
  \& {Pakmor}}]{2015MNRAS.453.3999M}
{Marinacci} F., {Vogelsberger} M., {Mocz} P., {Pakmor} R., 2015, \mnras, 453,
  3999

\bibitem[{{Martinsson} {et~al}\mbox{.}(2016){Martinsson}, {Verheijen},
  {Bershady}, {Westfall}, {Andersen}, \& {Swaters}}]{2016A&A...585A..99M}
{Martinsson} T.~P.~K., {Verheijen} M.~A.~W., {Bershady} M.~A., {Westfall}
  K.~B., {Andersen} D.~R., {Swaters} R.~A., 2016, \aap, 585, A99

\bibitem[{{Michel-Dansac} {et~al}\mbox{.}(2011){Michel-Dansac}, {Abadi},
  {Navarro}, \& {Steinmetz}}]{2011MNRAS.414L...1M}
{Michel-Dansac} L., {Abadi} M.~G., {Navarro} J.~F., {Steinmetz} M., 2011,
  \mnras, 414, L1

\bibitem[{{Mihos}(1999)}]{1999IAUS..186..205M}
{Mihos} J.~C., 1999, in IAU Symposium, Vol. 186, Galaxy Interactions at Low and
  High Redshift, {Barnes} J.~E., {Sanders} D.~B., eds., p. 205

\bibitem[{{Moetazedian} \& {Just}(2015)}]{2015arXiv150803580M}
{Moetazedian} R., {Just} A., 2015, ArXiv e-prints

\bibitem[{{Momany} {et~al}\mbox{.}(2006){Momany}, {Zaggia}, {Gilmore},
  {Piotto}, {Carraro}, {Bedin}, \& {de Angeli}}]{2006A&A...451..515M}
{Momany} Y., {Zaggia} S., {Gilmore} G., {Piotto} G., {Carraro} G., {Bedin}
  L.~R., {de Angeli} F., 2006, \aap, 451, 515

\bibitem[{{Morganson} {et~al}\mbox{.}(2016){Morganson}, {Conn}, {Rix}, {Bell},
  {Burgett}, {Chambers}, {Dolphin}, {Draper}, {Flewelling}, {Hodapp}, {Kaiser},
  {Magnier}, {Martin}, {Martinez-Delgado}, {Metcalfe}, {Schlafly}, {Slater},
  {Wainscoat}, \& {Waters}}]{2016arXiv160407501M}
{Morganson} E. {et~al.}, 2016, ArXiv e-prints

\bibitem[{{Naab} \& {Ostriker}(2006)}]{2006MNRAS.366..899N}
{Naab} T., {Ostriker} J.~P., 2006, \mnras, 366, 899

\bibitem[{{Newberg} {et~al}\mbox{.}(2002){Newberg}, {Yanny}, {Rockosi},
  {Grebel}, {Rix}, {Brinkmann}, {Csabai}, {Hennessy}, {Hindsley}, {Ibata},
  {Ivezi{\'c}}, {Lamb}, {Nash}, {Odenkirchen}, {Rave}, {Schneider}, {Smith},
  {Stolte}, \& {York}}]{new02}
{Newberg} H.~J. {et~al.}, 2002, \apj, 569, 245

\bibitem[{{Ogiya} \& {Burkert}(2016)}]{2016MNRAS.457.2164O}
{Ogiya} G., {Burkert} A., 2016, \mnras, 457, 2164

\bibitem[{{Ostriker} \& {Binney}(1989)}]{1989MNRAS.237..785O}
{Ostriker} E.~C., {Binney} J.~J., 1989, \mnras, 237, 785

\bibitem[{{Pakmor} \& {Springel}(2013)}]{2013MNRAS.432..176P}
{Pakmor} R., {Springel} V., 2013, \mnras, 432, 176

\bibitem[{{Pakmor} {et~al}\mbox{.}(2016){Pakmor}, {Springel}, {Bauer}, {Mocz},
  {Munoz}, {Ohlmann}, {Schaal}, \& {Zhu}}]{2016MNRAS.455.1134P}
{Pakmor} R., {Springel} V., {Bauer} A., {Mocz} P., {Munoz} D.~J., {Ohlmann}
  S.~T., {Schaal} K., {Zhu} C., 2016, \mnras, 455, 1134

\bibitem[{{Pe{\~n}arrubia} {et~al}\mbox{.}(2005){Pe{\~n}arrubia},
  {Mart{\'{\i}}nez-Delgado}, {Rix}, {G{\'o}mez-Flechoso}, {Munn}, {Newberg},
  {Bell}, {Yanny}, {Zucker}, \& {Grebel}}]{2005ApJ...626..128P}
{Pe{\~n}arrubia} J. {et~al.}, 2005, \apj, 626, 128

\bibitem[{{Pranav} \& {Jog}(2010)}]{2010MNRAS.406..576P}
{Pranav} P., {Jog} C.~J., 2010, \mnras, 406, 576

\bibitem[{{Price-Whelan} {et~al}\mbox{.}(2015){Price-Whelan}, {Johnston},
  {Sheffield}, {Laporte}, \& {Sesar}}]{2015MNRAS.452..676P}
{Price-Whelan} A.~M., {Johnston} K.~V., {Sheffield} A.~A., {Laporte} C.~F.~P.,
  {Sesar} B., 2015, \mnras, 452, 676

\bibitem[{{Purcell} {et~al}\mbox{.}(2011){Purcell}, {Bullock}, {Tollerud},
  {Rocha}, \& {Chakrabarti}}]{2011Natur.477..301P}
{Purcell} C.~W., {Bullock} J.~S., {Tollerud} E.~J., {Rocha} M., {Chakrabarti}
  S., 2011, \nat, 477, 301

\bibitem[{{Quillen} {et~al}\mbox{.}(2009){Quillen}, {Minchev},
  {Bland-Hawthorn}, \& {Haywood}}]{2009MNRAS.397.1599Q}
{Quillen} A.~C., {Minchev} I., {Bland-Hawthorn} J., {Haywood} M., 2009, \mnras,
  397, 1599

\bibitem[{{Quinn}, {Hernquist} \& {Fullagar}(1993){Quinn}, {Hernquist}, \&
  {Fullagar}}]{1993ApJ...403...74Q}
{Quinn} P.~J., {Hernquist} L., {Fullagar} D.~P., 1993, \apj, 403, 74

\bibitem[{{Radburn-Smith} {et~al}\mbox{.}(2014){Radburn-Smith}, {de Jong},
  {Streich}, {Bell}, {Dalcanton}, {Dolphin}, {Stilp}, {Monachesi}, {Holwerda},
  \& {Bailin}}]{2014ApJ...780..105R}
{Radburn-Smith} D.~J. {et~al.}, 2014, \apj, 780, 105

\bibitem[{{Reshetnikov} \& {Combes}(1998)}]{1998A&A...337....9R}
{Reshetnikov} V., {Combes} F., 1998, \aap, 337, 9

\bibitem[{{Reshetnikov} {et~al}\mbox{.}(2016){Reshetnikov}, {Mosenkov},
  {Moiseev}, {Kotov}, \& {Savchenko}}]{2016MNRAS.461.4233R}
{Reshetnikov} V.~P., {Mosenkov} A.~V., {Moiseev} A.~V., {Kotov} S.~S.,
  {Savchenko} S.~S., 2016, \mnras, 461, 4233

\bibitem[{{Ro{\v s}kar} {et~al}\mbox{.}(2010){Ro{\v s}kar}, {Debattista},
  {Brooks}, {Quinn}, {Brook}, {Governato}, {Dalcanton}, \&
  {Wadsley}}]{2010MNRAS.408..783R}
{Ro{\v s}kar} R., {Debattista} V.~P., {Brooks} A.~M., {Quinn} T.~R., {Brook}
  C.~B., {Governato} F., {Dalcanton} J.~J., {Wadsley} J., 2010, \mnras, 408,
  783

\bibitem[{{Sancisi} {et~al}\mbox{.}(2008){Sancisi}, {Fraternali}, {Oosterloo},
  \& {van der Hulst}}]{2008A&ARv..15..189S}
{Sancisi} R., {Fraternali} F., {Oosterloo} T., {van der Hulst} T., 2008, \aapr,
  15, 189

\bibitem[{{Schweizer} \& {Seitzer}(1992)}]{1992AJ....104.1039S}
{Schweizer} F., {Seitzer} P., 1992, \aj, 104, 1039

\bibitem[{{Sellwood}(2013)}]{2013pss5.book..923S}
{Sellwood} J.~A., 2013, {Dynamics of Disks and Warps}, {Oswalt} T.~D.,
  {Gilmore} G., eds., p. 923

\bibitem[{{Shen} \& {Sellwood}(2006)}]{2006MNRAS.370....2S}
{Shen} J., {Sellwood} J.~A., 2006, \mnras, 370, 2

\bibitem[{{Slater} {et~al}\mbox{.}(2014){Slater}, {Bell}, {Schlafly},
  {Morganson}, {Martin}, {Rix}, {Pe{\~n}arrubia}, {Bernard}, {Ferguson},
  {Martinez-Delgado}, {Wyse}, {Burgett}, {Chambers}, {Draper}, {Hodapp},
  {Kaiser}, {Magnier}, {Metcalfe}, {Price}, {Tonry}, {Wainscoat}, \&
  {Waters}}]{2014ApJ...791....9S}
{Slater} C.~T. {et~al.}, 2014, \apj, 791, 9

\bibitem[{{Sparke} \& {Casertano}(1988)}]{1988MNRAS.234..873S}
{Sparke} L.~S., {Casertano} S., 1988, \mnras, 234, 873

\bibitem[{{Springel}(2005)}]{springel2005a}
{Springel} V., 2005, \mnras, 364, 1105

\bibitem[{{Springel}(2010)}]{2010MNRAS.401..791S}
{Springel} V., 2010, \mnras, 401, 791

\bibitem[{{Springel} \& {Hernquist}(2005)}]{2005ApJ...622L...9S}
{Springel} V., {Hernquist} L., 2005, \apjl, 622, L9

\bibitem[{{Velazquez} \& {White}(1999)}]{1999MNRAS.304..254V}
{Velazquez} H., {White} S.~D.~M., 1999, \mnras, 304, 254

\bibitem[{{Vesperini} \& {Weinberg}(2000)}]{2000ApJ...534..598V}
{Vesperini} E., {Weinberg} M.~D., 2000, \apj, 534, 598

\bibitem[{{Villalobos} \& {Helmi}(2008)}]{2008MNRAS.391.1806V}
{Villalobos} {\'A}., {Helmi} A., 2008, \mnras, 391, 1806

\bibitem[{{Vogelsberger} {et~al}\mbox{.}(2013){Vogelsberger}, {Genel},
  {Sijacki}, {Torrey}, {Springel}, \& {Hernquist}}]{2013MNRAS.436.3031V}
{Vogelsberger} M., {Genel} S., {Sijacki} D., {Torrey} P., {Springel} V.,
  {Hernquist} L., 2013, \mnras, 436, 3031

\bibitem[{{Weinberg}(1998)}]{1998MNRAS.299..499W}
{Weinberg} M.~D., 1998, \mnras, 299, 499

\bibitem[{{Widrow} {et~al}\mbox{.}(2014){Widrow}, {Barber}, {Chequers}, \&
  {Cheng}}]{2014MNRAS.440.1971W}
{Widrow} L.~M., {Barber} J., {Chequers} M.~H., {Cheng} E., 2014, \mnras, 440,
  1971

\bibitem[{{Widrow} {et~al}\mbox{.}(2012){Widrow}, {Gardner}, {Yanny},
  {Dodelson}, \& {Chen}}]{2012ApJ...750L..41W}
{Widrow} L.~M., {Gardner} S., {Yanny} B., {Dodelson} S., {Chen} H.-Y., 2012,
  \apjl, 750, L41

\bibitem[{{Xu} {et~al}\mbox{.}(2015){Xu}, {Newberg}, {Carlin}, {Liu}, {Deng},
  {Li}, {Sch{\"o}nrich}, \& {Yanny}}]{2015ApJ...801..105X}
{Xu} Y., {Newberg} H.~J., {Carlin} J.~L., {Liu} C., {Deng} L., {Li} J.,
  {Sch{\"o}nrich} R., {Yanny} B., 2015, \apj, 801, 105

\bibitem[{{Yanny} \& {Gardner}(2013)}]{2013ApJ...777...91Y}
{Yanny} B., {Gardner} S., 2013, \apj, 777, 91

\bibitem[{{Yanny} {et~al}\mbox{.}(2003){Yanny}, {Newberg}, {Grebel}, {Kent},
  {Odenkirchen}, {Rockosi}, {Schlegel}, {Subbarao}, {Brinkmann}, {Fukugita},
  {Ivezic}, {Lamb}, {Schneider}, \& {York}}]{yanny03}
{Yanny} B. {et~al.}, 2003, \apj, 588, 824

\bibitem[{{York} {et~al}\mbox{.}(2000){York}, {Adelman}, {Anderson},
  {Anderson}, {Annis}, {Bahcall}, {Bakken}, {Barkhouser}, {Bastian}, {Berman},
  {Boroski}, {Bracker}, {Briegel}, {Briggs}, {Brinkmann}, {Brunner}, {Burles},
  {Carey}, {Carr}, {Castander}, {Chen}, {Colestock}, {Connolly}, {Crocker},
  {Csabai}, {Czarapata}, {Davis}, {Doi}, {Dombeck}, {Eisenstein}, {Ellman},
  {Elms}, {Evans}, {Fan}, {Federwitz}, {Fiscelli}, {Friedman}, {Frieman},
  {Fukugita}, {Gillespie}, {Gunn}, {Gurbani}, {de Haas}, {Haldeman}, {Harris},
  {Hayes}, {Heckman}, {Hennessy}, {Hindsley}, {Holm}, {Holmgren}, {Huang},
  {Hull}, {Husby}, {Ichikawa}, {Ichikawa}, {Ivezi{\'c}}, {Kent}, {Kim},
  {Kinney}, {Klaene}, {Kleinman}, {Kleinman}, {Knapp}, {Korienek}, {Kron},
  {Kunszt}, {Lamb}, {Lee}, {Leger}, {Limmongkol}, {Lindenmeyer}, {Long},
  {Loomis}, {Loveday}, {Lucinio}, {Lupton}, {MacKinnon}, {Mannery}, {Mantsch},
  {Margon}, {McGehee}, {McKay}, {Meiksin}, {Merelli}, {Monet}, {Munn},
  {Narayanan}, {Nash}, {Neilsen}, {Neswold}, {Newberg}, {Nichol}, {Nicinski},
  {Nonino}, {Okada}, {Okamura}, {Ostriker}, {Owen}, {Pauls}, {Peoples},
  {Peterson}, {Petravick}, {Pier}, {Pope}, {Pordes}, {Prosapio},
  {Rechenmacher}, {Quinn}, {Richards}, {Richmond}, {Rivetta}, {Rockosi},
  {Ruthmansdorfer}, {Sandford}, {Schlegel}, {Schneider}, {Sekiguchi}, {Sergey},
  {Shimasaku}, {Siegmund}, {Smee}, {Smith}, {Snedden}, {Stone}, {Stoughton},
  {Strauss}, {Stubbs}, {SubbaRao}, {Szalay}, {Szapudi}, {Szokoly}, {Thakar},
  {Tremonti}, {Tucker}, {Uomoto}, {Vanden Berk}, {Vogeley}, {Waddell}, {Wang},
  {Watanabe}, {Weinberg}, {Yanny}, {Yasuda}, \& {SDSS Collaboration}}]{sdss}
{York} D.~G. {et~al.}, 2000, \aj, 120, 1579

\bibitem[{{Younger} {et~al}\mbox{.}(2008){Younger}, {Besla}, {Cox},
  {Hernquist}, {Robertson}, \& {Willman}}]{2008ApJ...676L..21Y}
{Younger} J.~D., {Besla} G., {Cox} T.~J., {Hernquist} L., {Robertson} B.,
  {Willman} B., 2008, \apjl, 676, L21

\bibitem[{{Yurin} \& {Springel}(2014)}]{2014arXiv1411.3729Y}
{Yurin} D., {Springel} V., 2014, ArXiv e-prints

\end{thebibliography}

%%%%%%%%%%%%%%%%%%%%%%%%%%%%%%%%%%%%%%%%%%%%%%%%%%

%%%%%%%%%%%%%%%%%%%%%%%%%%%%%%%%%%%%%%%%%%%%%%%%%%

% Don't change these lines
%\bsp	% typesetting comment
\label{lastpage}
\end{document}